\documentclass[11pt]{article}
\usepackage[letterpaper, left=.65in, top=.65in, right=.65in, bottom=.65in,nohead,includefoot,ignoremp]{geometry}
\usepackage{times,bm,graphicx,natbib,amsmath,amssymb,amsfonts,amsthm,setspace,multirow,caption}
\usepackage[usenames,dvipsnames,x11names]{xcolor}
\usepackage[pdftex,pagebackref=true]{hyperref}
\usepackage{xy}\xyoption{all} \xyoption{poly} \xyoption{knot}  
\hypersetup{colorlinks,linkcolor=RoyalBlue1,urlcolor=RoyalBlue2, citecolor=RoyalBlue3}
\usepackage[center]{titlesec}
			\titlelabel{\thetitle.\quad}
			\titleformat{\subsection}{\normalfont\large\bfseries}{\thesubsection}{1em}{}
			\titleformat{\subsubsection}{\normalfont\em}{\thesubsubsection}{1em}{}
				\titlespacing*{\subsubsection}{\parindent}{1em}{1em}
\bibpunct{(}{)}{;}{a}{}{,}
\newcommand{\red}[1]{{\color{BrickRed}#1}}
\newcommand{\blu}[1]{{\color{RoyalBlue3}#1}}
\def\bigcdot{+}
\def\eqno#1{eqn.~(\ref{eq:#1})}
\def\ci{\perp\!\!\!\perp}

\begin{document}

\title{Scalable Bayesian Modeling, Monitoring and \\ Analysis of Dynamic Network Flow Data}
\author{
Xi Chen, 
Kaoru Irie, \\
David Banks,
Robert Haslinger, 
Jewell Thomas
\&
Mike West
}
\maketitle 

\begin{abstract} 
\noindent Traffic flow count data in networks arise in many applications, such as automobile or aviation transportation, 
certain directed social network contexts, and internet studies. Using an example of internet browser traffic 
flow through domains of an international news website, we present Bayesian analyses of two linked classes of models which, in tandem, allow fast, scalable and interpretable Bayesian inference.
We first develop flexible state-space models for streaming count data, able to adaptively
characterize and quantify network dynamics effectively and efficiently in real-time. We then use these
efficiently implemented models as {\em emulators} of more structured, time-varying gravity models that 
allow closer and formal dissection of network dynamics. This yields interpretable inferences on 
traffic flow characteristics, and on dynamics in interactions among network nodes. 
Bayesian model monitoring theory defines a strategy for sequential model
assessment and adaptation in cases of signaled departures of network flow data from 
model-based predictions.  Exploratory and sequential monitoring analyses of
evolving traffic on a defined network of web domains in e-commerce demonstrate the
utility of this coupled Bayesian emulation approach to analysis of streaming network count data.

\end{abstract}

\vfill
\vbox{  
\small
\noindent{\em \bf Author information:}
\begin{itemize} \itemsep-2pt 
\item Xi Chen  \href{mailto:xi@stat.duke.edu}{\it  $<$xi@stat.duke.edu$>$}
PhD student, Department of Statistical Science, Duke University.
\item Kaoru Irie  \href{mailto:iriekao@gmail.com}{\it  $<$iriekao@gmail.com$>$}
Assistant Professor, Department of Economics, University of Tokyo. \\
Kaoru's research on this project was performed 
while he was a PhD student in Statistical Science at Duke University.
\item David Banks \href{mailto:banks@stat.duke.edu}{\it  $<$banks@stat.duke.edu$>$} 
Professor,  Department of Statistical Science, Duke University.
\item Robert Haslinger 
\href{mailto:rob.haslinger@gmail.com}{\it  $<$rob.haslinger@gmail.com$>$} 
Lead Data Scientist, The Sync Project,  Boston. \\ Rob's research on this project was performed 
while he was Senior Data Scientist at MaxPoint Interactive Inc. 
\item Jewell Thomas \href{mailto:Jewell.Thomas@maxpoint.com}{\it $<$Jewell.Thomas@maxpoint.com$>$}
Staff Data Scientist, MaxPoint Interactive Inc. 
\item Mike West \href{mailto:mw@stat.duke.edu}{\it  $<$mw@stat.duke.edu$>$}
The Arts  \& Sciences Professor  of Statistics \& Decision
Sciences, Department of Statistical Science, Duke University.
\end{itemize} 
\noindent{\bf Acknowledgements:}

\noindent We are grateful to Mark Lowe of MaxPoint for discussion and input throughout the project that 
this paper represents. Support from the Nakajima Foundation (to Kaoru Irie) is gratefully acknowledged. 
Any opinions, findings and conclusions or recommendations expressed in this work are those of the authors and do not necessarily reflect the views of the Nakajima Foundation.

\medskip
\noindent{\bf Manuscript information:}
 
\noindent This  technical report is based on the earlier released Duke University Statistical Science 
		\href{http://stat.duke.edu/research/papers/2015-02}{Discussion Paper 2015-02},
		 July  2015 
}

\setcounter{page}0
\thispagestyle{empty}


\section{INTRODUCTION}
 
Increasing access to streaming data on dynamic networks drives interest in formal models 
to quantify stochasticity and structure of latent processes underlying observable data streams. 
Modeling interests are coupled with concerns to monitor and adapt to changing patterns, and 
to signal and highlight dynamics that may reflect interesting departures from the norm. 
Key challenges are real-time/sequential analysis and scalability:  interest lies in relevant statistical 
models whose analyses are inherently sequential in time, as well as computationally efficient and 
scalable with network size and sampling rates.  Relevant models should also define sound statistical 
methods for monitoring and short-term prediction, and elucidate the complexities and dynamics 
in network structure in both single sample inference and multi-sample comparisons across contexts. 

We contribute to this area with modeling and methodological developments coupled with a motivating
applied study of internet traffic in e-commerce. 
Consistent with the primary applied goals outlined above, the main contributions of this work are as follows.  
\begin{itemize}  \setlength\itemsep{0pt}
\item  A flexible and customized statistical modeling framework for: (i) characterizing patterns of
temporal variation in network flows at the levels of nodes and pairs of nodes;  
(ii) model-based exploratory data analyses of  network flows within and across contexts; and  
(iii) the ability to scale to large networks. 
\item  Use of these flexible, efficient models as Bayesian emulators of more structured 
network flow models. This yields 
computationally efficient dissection of the dynamics to evaluate node-specific and 
dependence/interaction effects across nodes in a structured model context where
analysis is otherwise computationally challenging in more than small networks. 
\item  Formal Bayesian model assessment methodology for sequential monitoring of flow
patterns with the ability to signal departures from predictions in real-time and allow informed
interventions as a response, and in a scalable framework.
\item Development and validation of the above in exploratory and monitoring analyses of 
data from the motivating application; here the observations are  
streaming counts of visitors in a set of 
defined web domains  (collections of webpages) in a structurally well-defined but dynamic/evolving 
website. This includes evaluation of node-specific and node-pair interactions in the flow dynamics within the network over a given time period,  comparisons across time periods and across days, and analyses utilizing Bayesian monitoring and adaptation to respond to departures from predicted flow patterns.
\end{itemize} 
Following a discussion of the motivating applied setting, network and data in Section~\ref{sec:webdata}, 
we develop our class of Bayesian dynamic flow models (BDFMs) in Section~\ref{sec:BDFM}. 
BDFMs are flexible univariate dynamic models for series of counts representing 
flows into the network and between within-network node pairs. These (non-stationary and 
non-normal) state-space models  for streaming count data rely on discrete-time
gamma processes historically used in volatility modeling, and that have very 
recently become of interest as flexible smoothing and short-term predictive models for 
space-time processes underlying count data. Our use of these models for within-network flows 
is novel and involves 
methodological extension to adapt and customize them to provide suitable univariate emulators of the 
underlying, inherent dynamic multinomial structures governing flows at each time point. 
This use of sets of {\em decoupled} univariate models that are then {\em recoupled} 
to define the actual multinomial probability processes is: (a) explicitly designed to be 
computationally efficient in  on-line data analysis, scaling quadratically in the number of 
network nodes and  enabling distributed implementation for streaming data 
on large networks; (b) allows for diverse patterns in the dynamics of flow rates that 
a time-varying Dirichlet-multinomial model simply annot; and (c) relates to the recent 
development of conceptually similar (decouple/recouple) approaches that have advanced 
multivariate dynamic modeling in conditionally normal contexts~\citep{GruberWest2015BA,ZhaoXieWest2015,GruberWest2016portfolios}. 
Section~\ref{sec:BDFMdata} discusses some aspects and summaries of exploratory analysis 
of the network flow data from a defined MaxPoint network of the Fox News website. 
This highlights the use of customized BDFMs, with one focus on 
exploring aspects of flow dynamics on the network across the same time periods on 
different days.  The supporting appendix material gives additional technical details and
discussion. 

Section~\ref{sec:DGM} introduces a class of more highly structured {\em dynamic 
gravity models (DGMs)} for network flows. These are non-normal, log-linear random-effects models
with time-varying parameters for flow rate contributions of 
origin nodes, destination nodes and origin-destination interaction effects.
Our DGMs extend prior work with static gravity models~\citep[e.g.][]{West1994, Sen:1995,Cogdon2000,JandarovEtAl2014} 
to the time-varying parameter context,  defining a class of models able to represent 
complicated patterns of dependency structure, and their temporal variations, across nodes. 
Importantly, we show that the flexible and computationally simple BDFM framework can 
be mapped one:one to that of the DGM.   This underlies one further novel contribution of this 
work: the  use of the fast, efficient BDFMs as {\em emulators} of DGMs. This is key
from the viewpoint of scalability; fitting gravity models, even without time-varying
parameters, is a challenging issue in more than modest dimensional networks, and simply 
infeasible  in any realistic dynamic extension appropriate for scalable, on-line 
analysis of streaming network flow data.  Further, we avoid the challenging approach of 
defining and parametrizing time-evolution models for DGMs directly,  adopting the implicit 
structures induced in the mapping from BDFMs where model specification and fitting is relatively facile.  
Example results and highlights from the Bayesian emulation analysis of DGMs for 
the MaxPoint Fox News study appear in 
Section~\ref{sec:DGMdata}. 

Section~\ref{sec:monitor} develops methods of formal, sequential Bayesian model monitoring and
adaptation (automatic intervention)  for BDFMs.  The aim here is to build into the fast, 
decoupled analysis an ability to efficiently evaluate incoming flows against model predictions
so as to signal data-- at the level of individual nodes and node-pairs-- that appear discrepant, 
and that may signal outliers or changes in flow trends/rates beyond the norm.  In addition to 
signaling such events and thus providing opportunity for direct intervention, we couple monitoring 
with the use of automatic intervention to allow the model to appropriately
adapt to data at the next few time points.   This Bayesian testing/adaptation strategy builds on
core theory underlying its use in time series forecasting contexts with dynamic linear models~(\citealp{West1986a,West1986,Harrison1987,West1989}; see also chapter 11 of~\citealp{WestHarrison1997}).   Some of the novelty here is in the use of these 
ideas in non-linear, non-normal dynamic models for count data-- our class of BDFMs. 
Importantly, monitoring is applied in parallel across nodes and node-pairs, so is also 
scalable with network size.  Some departures from normal variation in patterns of
flow may be related across nodes, and the approach has an ability to explore and evaluate this
both within the BDFM model context and then following the map to more structured DGMs 
that directly reflect interaction effects. 
Application to the Fox News network data highlights
some aspects of the use of this in connection with selected  network nodes. 
Summary comments conclude the paper in Section~\ref{sec:closingcomments},
and additional supporting material is given in the Appendix.

\section{Web-Domain Network Flows \label{sec:webdata} } 

\subsection{Context and Data \label{sec:contextdata} }

Our context is traffic flow among {\em domains} (defined sets of pages) of the Fox News website.
Domains include the Homepage and several categories of news and consumer content-- such as 
Politics, Entertainment, Travel, Science, etc.-- as defined by Fox News. 
While the domain structure is persistent, the nature of webpage definition and content within a domain
is dynamic; content changes on a daily basis (updated at midnight) 
but also more rapidly when  noteworthy events occur. 
MaxPoint places ads on pages in these Fox News domains, and thus can 
track flows of anonymized users as they move through its pages. 
While some users can be tracked individually, this is not the norm, and we focus in this paper
on aggregated flow counts, not the trajectories of individuals.

On-line advertisers are interested in a host of  statistical issues related to traffic flow and domain content.
The field has become quite sophisticated, employing complex  recommender systems~\citep{Koren:2009}, 
sentiment analysis~\citep{Pang:2008}, text mining \citep{Soriano:2013}, and other 
methods~\citep{Agarwal:2010, Taddy:2013}. However, basic questions of understanding and characterizing 
traffic  across domains have not received the attention they require. In particular, there is commercial value 
in identifying how the popularity of a site changes on short time scales, and how sites interact with respect to 
traffic. As example, our data showed a morning-after spike in traffic to the Entertainment domain following 
the Grammy awards, which would have been an opportunity to market concert tickets; unusual interactions 
in flows between Science and Health may reflect new medical findings that might incline people to 
purchase gym memberships;  increased flow rates from Homepage to Science that contradict the
general stable or somewhat decreasing trends of overall traffic may indicate specific opportunities to target
scientific product consumers. 

As pages within a domain are updated, questions arise as to whether browsing traffic patterns change as 
a result.  To address this statistically, we need to understand  stochastic variation in past browser traffic 
so that comparisons can be made of incoming traffic streams against recent statistical \lq\lq norms", and 
significant deviations from short-term predictions based on current dynamic patterns can be identified. 
Companies that have flow models which enable them to predict how traffic will change as content changes, 
that are able to sensitively characterize and monitor patterns of change in interactions as well as overall rates, 
and that can signal anomalous changes to provide opportunities for intervention and actions,  will be advantaged.
They can recognize opportunities more quickly, and-- for example-- may then adapt bidding strategies for relevant  keywords to be dynamically calibrated to  expected revenue.

\subsection{Data} 

The data set 
contains Fox News website 
visit data during 09{:}00--10{:}00am and
01{:}00--02{:}00pm EST on each of six days, 
February 23rd-24th, March 2nd-3rd and 9th-10th, 2015.
These days are Mondays or Tuesdays. 
%
Since the Fox News website structure changes often, with new pages
being added and old pages being archived, the analysis aggregates  webpages into groups specified by the host domain
{\tt www.foxnews.com},
 and the set of first url paths after the host domain,  including
 examples such as 
e.g. {\tt www.foxnews.com/politics/*}
and {\tt www.foxnews.com/US/*}.
These classify all pages into 22 
domains: Homepage, Politics, US, Opinion, Entertainment, Technology, 
Science, Health, Travel, Leisure, World News, Sports, Shows, Weather, 
Category, Latino, Story, On-Air, Video, National News, Magazine, and Other. 
 
The data set includes anonymized visitors from nearly every time zone on the planet.
In order to study time-of-day effects, such as, say, a tendency to browse
news in the morning and entertainment in the afternoon, it is necessary
to stratify by time zone. Here we focus  on users in the Eastern North
America time zone; those are the most numerous, and the two time
windows used in this study were chosen with the expectation that 
different browsing patterns might occur at those times.

Aggregate data give time series of counts in half-minute intervals, i.e., 
$120$ time points of domain occupancy,  flows from each domain, and 
flows into each domain. 
In each half minute interval, if the record shows the same user in
two or more domains, then each of her/his moves is counted in the flow
data into each of these domains.  
If the user refreshes the same page multiple times spanning more than one time interval, 
then s/he is counted as simply staying in that domain; this can be done as 
the web browsing tool performs automatic refresh. 
Importantly, if a user stays in the same domain for more than five 
minutes,  s/he is declared as no longer active and counted as 
leaving the Fox News site.
If such a user later appears in one domain, s/he counts as inflow
from outside the Fox News site.
Finally, we cannot track user information either before or
after the one-hour observation window; we thus restrict attention to the period 09{:}05--09{:}55am 
and  01{:}05--01{:}55pm, consisting of uncensored flows, 
using the first 5 minutes of data informally to define priors. 
Thus the series runs from $t=1{:}T$ with $T=110$ in each time period.

Aggregation at half-minute intervals reflects a balance of 
interests in fine-time scale modeling against information content of data with low 
flow rates.    For domain
pairs with low flow rates it is sometimes too low, leading to excessive volatility
in the BDFM and noisy parameter estimates in the DGM.  No single window 
is good for all node pairs at all times, but preliminary exploration found this to 
be a good compromise.
The decision that a user is inactive after five minutes is based 
on previous research on how users access on-line articles.
Few people read more than the first paragraph of a news story, and at
the times of day for which data were collected, interruptions are likely.
Investigations of on-line session length have focused on dynamics of search engine 
use~\citep[e.g.][and references therein]{Silverstein1999,Qiu2005} in settings where the full breadth 
of user browsing behavior is visible. 
Using additional context such as change in search engine query topic, these studies have 
derived average user session lengths between 5--60minutes~\citep{Jansen2007}. We employed 
a session length limit on the lower end of this spectrum due to the fact that a user is likely to 
only spend a fraction of a larger total browsing session on Fox News.

\subsection{Network Structure and Notation} 

\begin{figure}[htbp!]
\centering
  \includegraphics[width=5in]{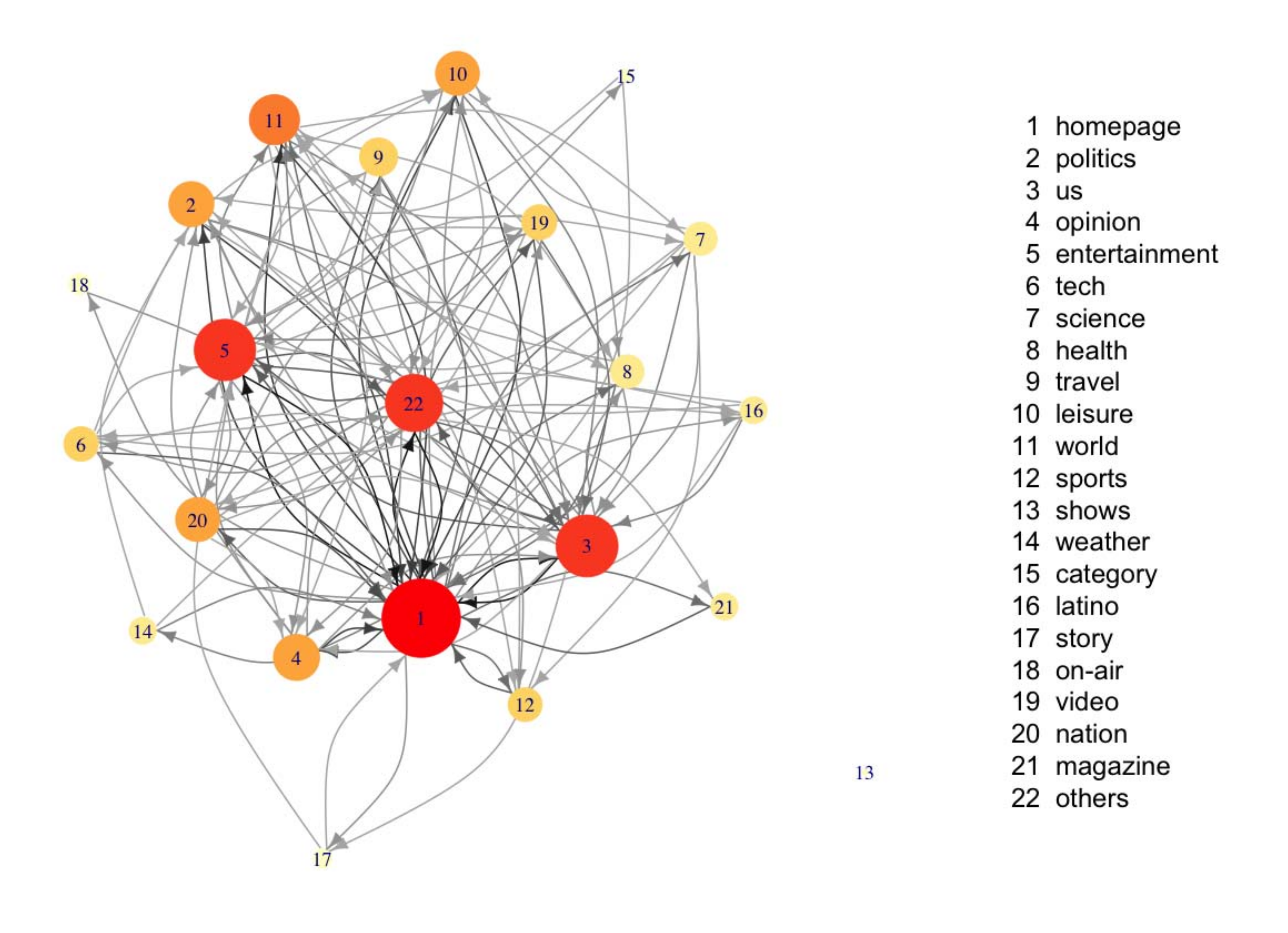} 
\caption{A snapshot of counts and flows on the Fox News network at 
time time $t=1$ (09{:}05{.}30 on February 23rd 2015).
The circular numbered nodes represent the domains, with 
diameters  (and, in the online paper, color) 
proportional to occupancy $n_{it}$ for node $i$ at this time point $t.$ 
Each arrow $i\rightarrow j$ has width proportional to flow $x_{ijt}.$  }
\label{fig:networkmovie}
$$ 
\xymatrix{ 
\textrm{Inflow} \ar@/_0pc/_*+{x_{0it}}[rrrd]  & & & & & & *+++[o][F-]{1} \ar@{.}[d] & \\
& & &
	*+++[o][F-]{i} \ar@/^/@{-->}^*+{x_{i1t}}[rrru] 
	\ar@{-->}^*+{x_{ijt}}[rrr] \ar@/_/@{-->}^*+{x_{iIt}}[rrrd] 
	\ar@/_0pc/@{-->}_*+{x_{i0t}}[llld] & & & *+++[o][F-]{j} \ar@{.}[d] & \\
 \textrm{Outflow} & & & & & & *+++[o][F-]{I} & \\
}
$$ 
\caption{Network schematic and notation for flows at time $t$.}
\label{fig:networkcartoon}
\end{figure}

Referring to sites external to the Fox News website as node 0,  we have
$23$ network nodes; the $I=22$ actual domains and ``External", indexed 
as $i=0{:}I.$    At each time $t=1{:}T,$ define 
$x_{ijt}$ as the flow count from node $i$ to $j$, 
including the inflows $x_{0it}$ and outflows $x_{i0t}$ relative to 
the External domain.  Also, denote the 
number of occupants of node $i$ at the end of the $t$ period by
$n_{it}$-- a random quantity at the start of the period, but then known and
given by the sum of inflows minus outflows at the end of the period. 
  Figure~\ref{fig:networkmovie} 
provides a visualization of data at the first time interval, and the 
schematic in Figure~\ref{fig:networkcartoon} reflects our notation. 

If $\phi>0$ has a gamma distribution with shape $r$ and scale $1/c$, we 
write $\phi \sim Ga(r,c)$, noting that $E(\phi)=r/c$ and $V(\phi)=r/c^2. $ 
If $\eta\in (0,1)$ has a beta distribution with p.d.f. proportional to $\eta^{a-1}(1-\eta)^{b-1}$
for $a>0,b>0,$ we write $\eta \sim Be(a,b).$
If the $n-$vector of counts $z$ has a multinomial distribution with  total counts $n$
and probability vector $\theta,$ we write $z\sim Mn(n,\theta).$ For any 
series of random quantities $z_t$ over $t=0,1,\ldots,$ we use the succinct notation
$x_{h:k} = \{ x_h, x_{h+1}, \ldots, x_k \}$ for any indices $h\le k.$

\bigbreak

\section{Fast, Flexible Dynamic Flow Models \label{sec:BDFM} }

\subsection{Background: Discrete-Time Volatility Models for Poisson Rates \label{sec:gammabetadm}} 

BDFMs are based on gamma-beta random walks that 
have been key to stochastic volatility modeling 
 for more than 30 years~\citep[][Section 10.8]{West1989book,WestHarrison1997}. 
Based initially on Bayesian discount concepts related to exponential smoothing
for volatilities~\citep{Ameen1985,Harrison1987,Quintana1987} and with \lq\lq steady" evolutions in
non-Gaussian dynamic models~\citep{Smith1979}, these {\em gamma-beta discount models}
yield closed form Bayesian analyses.  They have seen some-- though limited-- use as  
models for rates underling time series of conditionally Poisson counts~\citep{Harvey1989,Brandt2000},
which is a starting point here.  We extend the applicability
of the basic model in a number of ways, with novel model forms customized to the
network flow context and that define flexible models for conditional multinomial data with
time-varying probabilities that go beyond prior use. We also heavily utilize
 full Bayesian posterior simulation of posteriors for latent rate processes,  
extending the use from the normal/linear dynamic volatility 
modeling context~\citep[][Chapter 4]{PradoWest2010}.

The essentials of gamma-beta discount models follow;  see Appendix~\ref{app:betagammadm}
for further details and discussion. In generic notation, suppose that $x_t$ is
a time series with $x_t|\phi_t \sim Poi( m_t\phi_t)$ conditionally independently  over $t=1,2,\ldots. $ 
Here $\phi_t$ is a latent level process and 
$m_t$ a scaling factor known at time $t$.  The $\phi_t$ process 
evolves via the Markov model 
\begin{equation}\label{eq:phieta}
\phi_{t}=\phi_{t-1} \eta_{t}/\delta_t, \qquad \eta_t \sim Be(\delta_t r_t, (1-\delta_t)r_t), \qquad
\eta_t \ci \eta_s, \phi_s, \,\, s<t,
\end{equation}
where  $\delta_t\in (0,1)$ is a specified {\em discount factor}  and 
$r_t$ is a known function of $t,x_{0{:}t-1},$ and 
independent innovations $\eta_t/\delta_t$ drive the evolution. The beta distributions imply:
 (i)  $E(\phi_{t}|\phi_{t-1}) = \phi_{t-1},$  hence this is a multiplicative random walk model, 
or ``steady" evolution; (ii)  a lower value of $\delta_t$ leads to a 
more diffuse distribution for $\eta_t/\delta_t$, and hence increased uncertainty about $\phi_t$ 
and adaptability to changing rates over time; 
a value closer to one indicates a steady, stable evolution. 

The model is structured to ensure full conjugacy in the forward filtering/Bayesian
sequential learning over time, and in retrospective analysis.   This is reflected in some key
summaries, as follows and with further details in Appendix~\ref{app:betagammadm}. 
Here $x_0$ is a synthetic notation for  initial  information. 
\begin{itemize}  \setlength\itemsep{0pt}
\item 
{\em Forward Filtering (FF):}
At any time $t,$ both the prior $p(\phi_t|x_{0{:}t-1})$ and the 
posterior  $p(\phi_t|x_{0{:}t})$ for the \lq\lq current" latent level are gamma distributions, with
trivially computed parameters that are updated as $t$ evolves
\item 
{\em One-Step Forecasts:} 
The one-step ahead forecast distribution made at time $t-1$ to predict time $t$  is 
generalized negative binomial with p.d.f. in \eqno{onestepnegbinpdf}.  On observing $x_t,$ 
the p.d.f. is trivially evaluated  to feed into computation of model 
marginal likelihoods (MMLs, as in Appendix) for assessment.
\item 
{\em Backward Sampling:} 
At end time $T,$  recursive simulation generates {\em time trajectories} 
$\phi_{0{:}T}$ of the rate process under its full posterior
$p(\phi_{0{:}T}|x_{0{:}T}).$   The computations are trivial, as detailed in
Section~\ref{sec:bs}.
\end{itemize} 
The model can be defined by
any sequence of specified discount factors $\delta_t.$  A constant value over
time defines a global smoothing rate; values closer to 1 constrain the stochastic innovation 
and hence the change from $\phi_{t-1}$ to $\phi_t$; smaller discount factor values 
lead to greater random changes in these Poisson levels.   Intervention to specify 
smaller discount factors at some time points, to reflect or anticipate higher levels of 
dynamic variation at those times, are sometimes relevant.   In our network flow models
below, we customize the specification of the sequence of discount factor to address issues that
arise in cases of low flow levels.  That extension of discount-based modeling defines the $\delta_t$ 
as time-varying functions of an underlying base discount rate, and the latter 
are then evaluated using MML measures. 

This   model provides the basis for 
flows into network nodes; we adapt and generalize it to define components
of flexible multinomial dynamic models for flows between nodes in a network. 
 
\subsection{Network Inflows: Poisson  Dynamic Models \label{sec:inflows} }

With notation for inflows as in Figure~\ref{fig:networkcartoon},  we adopt the
general model of Section~\ref{sec:gammabetadm} by adding suffices $i$ for network nodes and
setting the Poisson mean scaling factors to 1.  We now customize this model via
specification of discount factor sequences. At any node $i,$  the time $t$ 
inflow to node $i$ is $x_{0it} \sim Poi(\phi_{it})$
independently across  nodes $i=1{:}I,$ and  the latent 
levels $\phi_{it}$ follow node-specific gamma-beta discount models with discount factor
$\delta_{it}$ at time $t.$   The time $t \to t+1$ update/evolve 
steps are: (i) the time $t$ prior 
	$ \phi_{it}|x_{0i,0{:}t-1}  \sim Ga(\delta_{it} r_{i,t-1}, \delta_{it} c_{i,t-1}) $
	updates to the posterior
	$\phi_{it}|x_{0i,0{:}t} \sim Ga(r_{it}, c_{it})$ 
with $r_{it}=\delta_{it} r_{i,t-1} + x_{0it}$ and $c_{it}=\delta_{it} c_{i,t-1}+1.$ 
This then evolves to the time $t+1$ prior
	$ \phi_{i,t+}|x_{0i,0{:}t} \sim Ga(\delta_{i,t+1} r_{i,t},\delta_{i,t+1} c_{it})$, 
and so on. Specifying discount factors $\delta_{it}$ relates to the information 
content of gamma distributions as measured by the shape parameters $r_{i*}$; 
evolution each time point reduces this by discount factor, the latter representing a 
per-time-step decay of information induced by the stochastic evolution.  Our specification of 
discount rates is motivated by the following considerations. First, baseline levels
of variation on $\phi_{it}$ are likely to be node specific, so that each node should have its
own baseline discount rate to be assessed in data analysis. Second, 
 in cases of zero flow rates for a period of time,  $r_{it}$ is continually discounted and 
shrinks towards 0 while $c_{it}$ is incremented by 1 at each update step. That is, 
discounting is not  balanced by the prior-posterior update and the generates 
more and more diffuse posteriors favoring lower and lower $\phi_{it}.$ Ideally,
the posterior and prior should be very similar in cases of 0 flows,  and we address this 
with the specification 
 $\delta_{it} = d_i + (1-d_i)\exp(-k r_{i,t-1})$ at each $i,t,$ where $d_i$ is a 
{\em constant baseline discount factor} for node $i$  and $k>0$ a specified constant. 
The aim is that $\delta_{it}$ be close to the baseline unless information content   is very low; this our applied studies take $k=1$ (so that $\delta_{it}$ be close to-- within 10\% of-- the baseline
when $r_{i,t-1}>2$).
Then in cases of high information content,  the effective $\delta_{it}$ is  close to $d_i;$ 
otherwise, $\delta_{it}$ will be closer to 1 in cases of low information content, so 
appropriately limiting the decay of information in such cases.  

Node-specific MML measures that feed into model assessment
to aid in selection of the baseline discount factors $d_i.$     These measures of 
short-term predictive fit of the models can also be 
monitored sequentially  over time for on-line tracking of model performance. 
This ability to flag anomalous data at one node or any 
subset of nodes is key to commercial application of the
analysis, since that corresponds to new opportunities or
new threats (e.g., offer concert tickets after the Grammy Awards, but 
not on David Bowie's obituary). This view on anomaly detection 
is extended below, in Section~\ref{sec:monitor}, using  Bayesian model monitoring 
concepts.  One aspect of this is the ability to signal a need to {\em temporarily} reduce 
the value of the discount factor for a node at a time of degradation of
predictive performance that may relate to changes in $\phi_{it}$ that are 
larger than the \lq\lq standard" baseline discount factor value $d_i$ determines.

\subsection{Transitions from Network Nodes: Multinomial  Dynamic Models \label{sec:transits}} 

Transitions from any node $i$ at time $t$ are inherently multinomial with time-varying 
transition probabilities. To build flexible and scalable models for dynamics and dependencies in 
transition probability vectors is a challenge, with computational issues for even simple models
quickly dominating. Novel models here adapt and extend the univariate Poisson/gamma-beta 
random walk models to enable flexibility in modeling node-pair specific 
effects as they vary over time as well as scalability. 

Considering flows from node $i$ to node $j$ at time $t,$ and using notation
as in Figure~\ref{fig:networkcartoon}, the core model is
$x_{i,0{:}I,t} \sim Mn(n_{i,t-1},\theta_{i,0{:}I,t})$ where the 
current node $i$ occupancy level is $n_{i,t-1},$  and 
$\theta_{i,0{:}I,t}$ is the $(I+1)$-vector of transition probabilities $\theta_{ijt}$ 
(including the \lq\lq external" node-- i.e., leaving the Fox News network-- at $j=0).$
We structure {\em decoupled} BDFMs in terms of positive flow rates $\phi_{ijt}$ underlying each $x_{ijt}.$ Specifically, 
\begin{equation} \label{eq:mnflows} 
 x_{ijt} \sim Poi( m_{it}\phi_{ijt} ), \qquad m_{it} = n_{i,t-1}/n_{i,t-2},
\end{equation}
independently, with independent  gamma-beta evolutions for each  latent level $\phi_{ijt}.$  These BDFMs for each node pair 
can be customized with node-pair specific discount factors, allowing greater or lesser degrees
of variation by node pair. The set of models for elements of  $\phi_{i,0{:}I,t}$ implies 
a dynamic model for the vector of transition probabilities $\theta_{i,0{:}I,t}$ having elements
$\theta_{ijt}=\phi_{ijt}/\sum_{j=0{:}I} \phi_{ijt}.$   
Independence across nodes enables scaling, as the 
analyses can then be decoupled and run in parallel for the $\phi_{ijt}$ and then recoupled
to infer the $\theta_{ijt}.$ Dependencies in patterns of changes in the
$\phi_{ijt}$ are recovered in evaluating the posterior distributions and, as in Section~\ref{sec:DGM}, in using this set of models to emulate  gravity models that explicitly characterize  interdependencies.
 
A key and critical component of the model is the definition of the scaling factors $m_{it}$
in~\eqno{mnflows}.  
In decoupling the multinomial flows from node $i$ into parallel Poisson models for nodes $i\to j$ 
the inherent dependency on total occupancy of node $i$ is lost.  We restore this in using this
specific definition of  scaling factors to explicitly correct for occupancy changes.  
This recognizes that the decoupled, scaled models are  not predictive of overall occupancy-- rather, 
they are  decoupled, tractable models that are relevant to tracking and short-term prediction of 
{\em relative}  occupancy levels through the implied multinomial probabilities. The relevance of this scaling factor is most evident in cases of major changes in occupancy, when an abrupt increase in 
 node $i$ occupancy $n_{i,t-1}$ at time $t-1$ relative to its prior value $n_{i,t-2}$ will
lead to increased flows to other nodes at time $t$ even if the underlying transition probabilities
$\theta_{i,0{:}I,t}$ are essentially constant.   In such a case, the scaling factor will encourage 
the appropriate view that the $\phi_{ijt}$ are stable. Then, inferences on the $\phi_{ijt}$ directly 
yield inferences on the transition probabilities of interest: theoretically, the conditional 
multinomial probabilities are simply not impacted by the scaling factors, i.e., 
$$\theta_{ijt}=m_{it}\phi_{ijt}/\sum_{j=0{:}I} m_{it}\phi_{ijt} =
\phi_{ijt}/\sum_{j=0{:}I} \phi_{ijt}.$$ 
The theory and analysis details of Section~\ref{sec:gammabetadm} 
and Appendix~\ref{app:betagammadm} now apply with data and latent flow levels 
indexed by origin node $i$ and destination node $j.$   As with inflow models, 
we have flexibility to choose discount factors specific to context.
Following the discussion of Section~\ref{sec:inflows}, we specify
 $\delta_{ijt} = d_{ij} + (1-d_{ij})\exp(-k r_{ij,t-1})$ at each $i,j,t,$ where $d_{ij}$ is a 
{\em constant baseline discount factor} for node pair $i,j$  and $k>0$ a specified constant. 
Again this is later overlaid with intervention to adjust discount factor values as needed, 
based on sequential monitoring of flow patterns and using the MML measures--
now of course for each node pair-- as one formal  guide to model adequacy. 

In sequential analysis of transitions, the node-pair specific models generate full joint
predictions one-step ahead (or more, if desired) for the theoretically exact set of multivariate
flow vectors $x_{i,0{:}I,t}$ across all nodes.   The one-step forecast 
distribution does not have an analytic closed form,  but is trivially simulated to define forecasts. 
That is: (i)  simulate directly from each of the gamma-beta evolutions for the $\phi_{ijt};$
(ii) transform sampled values to the conditional multinomial 
 probabilities $\phi_{ijt}$; then (iii) sample the multinomial 
$x_{i,0{:}I,t} \sim Mn(n_{i,t-1},\theta_{i,0{:}I,t})$ at these parameter values.
Similarly, for both on-line and retrospective inference about transition probabilities,  samples 
from posteriors for the $\phi_{ijt}$ again simply transform to the required probability scale.

\subsection{Aspects of BDFM Analysis of Fox News Data \label{sec:BDFMdata}}

The analysis was applied separately to data from each of the six days.  
We focus on the am period of February 23rd 2015 for initial summaries, and then 
make some comparisons across days. 
Priors for the inflow rates are  
$\phi_{i0}|x_{0i0} \sim Ga(r_{i0},c_{i0})$ with $c_{i0}=1$ and $r_{i0}=c_{i0} z_i$
where $z_i$ is from inflows in the 5minutes {\em prior} to the start of model analysis at $t=1.$  
The priors for the node-node flows are, similarly,
$\phi_{ij0}|x_{0ij0} \sim Ga(r_{ij0},c_{ij0})$ with $c_{ij0}=1$ and with 
$r_{ij0}=c_{ij0} z_{ij}$
where each $z_{ij}$ is a point estimate based on that prior period. 
With $c_{*}=1$ the priors are relatively diffuse, and for most nodes one or a few initial 
observations ``wash out'' the effect of the prior. While some node-node pairs have 
very low counts, they all see traffic that then updates shape parameters over 
a few early periods;   some network links have counts in the thousands, 
while the average is around 40--50 across the time period. 

\begin{figure}[htbp!]
    \centering 
     \includegraphics[width=4.5in]{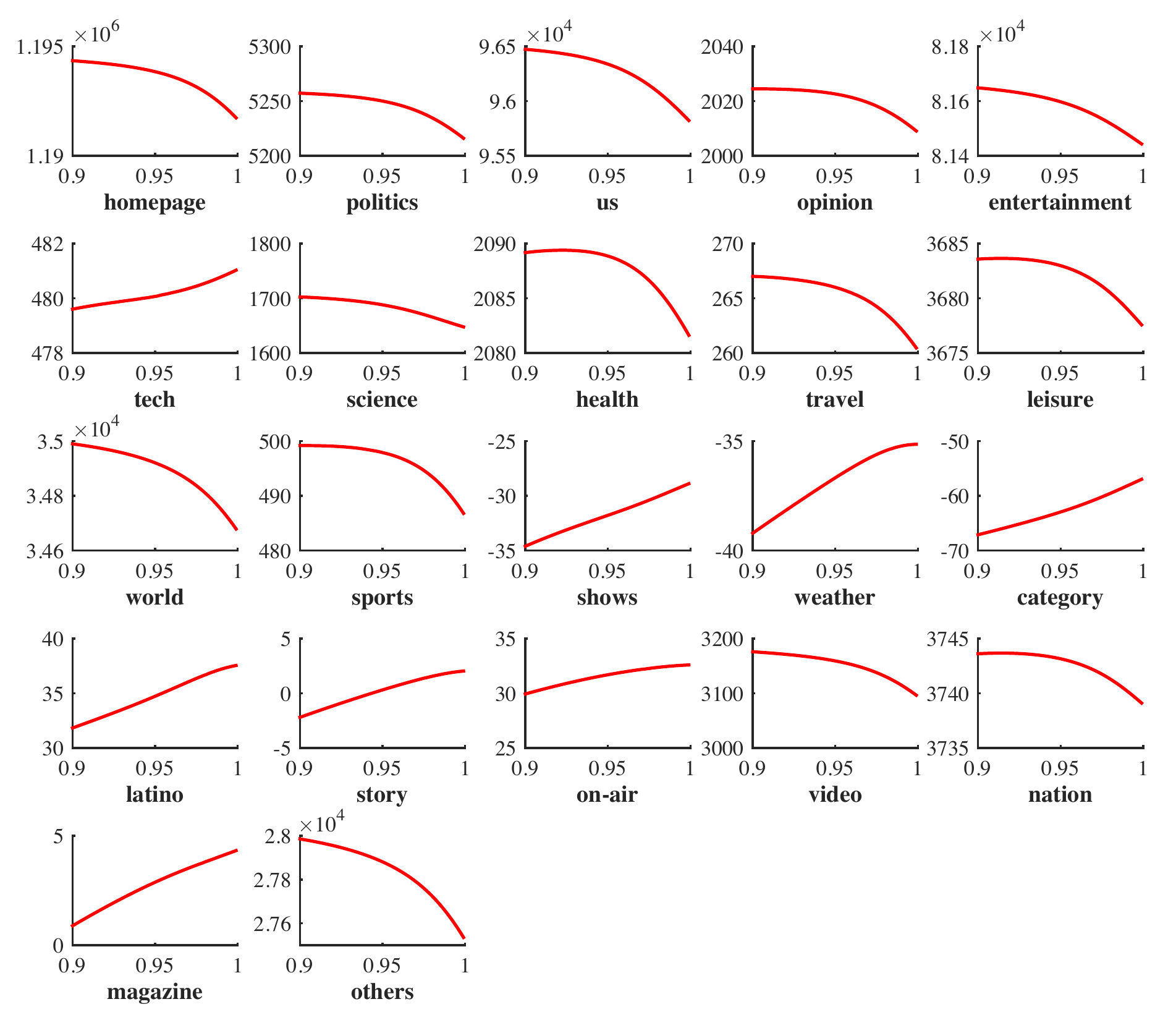} 
  \caption{ 
\small 
Marginal log-posteriors for baseline discount factors $d_i$ for the inflows
to Fox News nodes $i=1{:}22$ for the February 23rd 2015 am period.  
  }
  \label{fig:dfall}
\end{figure}

The prior for each baseline discount factor is a smoothness encouraging $Be(19,1)$ 
truncated to (0.9,0.999); reanalysis using uniform priors on this range  
led to little in the way of noticeable differences, as the marginal likelihoods at 
$t=T$  dominate.   
Running models in parallel across discrete grids of each discount
factors and evaluating MML measures at $t=T$ gives marginal likelihoods
that are mapped to posteriors.
Figure~\ref{fig:dfall} plots posteriors for the $d_i$ in the inflow models.  Some nodes exhibit higher volatility in flows  than others, consistent with smaller 
discount factors; these are particularly associated with   domains with high flow counts (e.g., inflows 
to Homepage).   Constraining the range to higher values is 
consistent with the expectations to generally \lq\lq smooth" trajectories for the $\phi_{\ast t}$ 
processes, which turns out to be consistent with the majority of flows; allowing smaller values
has little or no impact on much of the reported analysis. However, for some node flows
with patterns of higher levels of change and variation, lower discount factors would
lead to posteriors that suggest more volatile trajectories in a few cases. These are better
addressed in a model that uses a higher discount factor as standard,  but then with interventions to
allow increased uncertainty and adaptation in response to discrepant observations (whether
single or in batches); our developments in Section~\ref{sec:monitor} are heavily motivated by this.  The model is, of course, open to specification of whatever priors a user may regard as relevant or wish to explore.

\begin{figure}[htbp!]
    \centering 
    \includegraphics[width=3.8in]{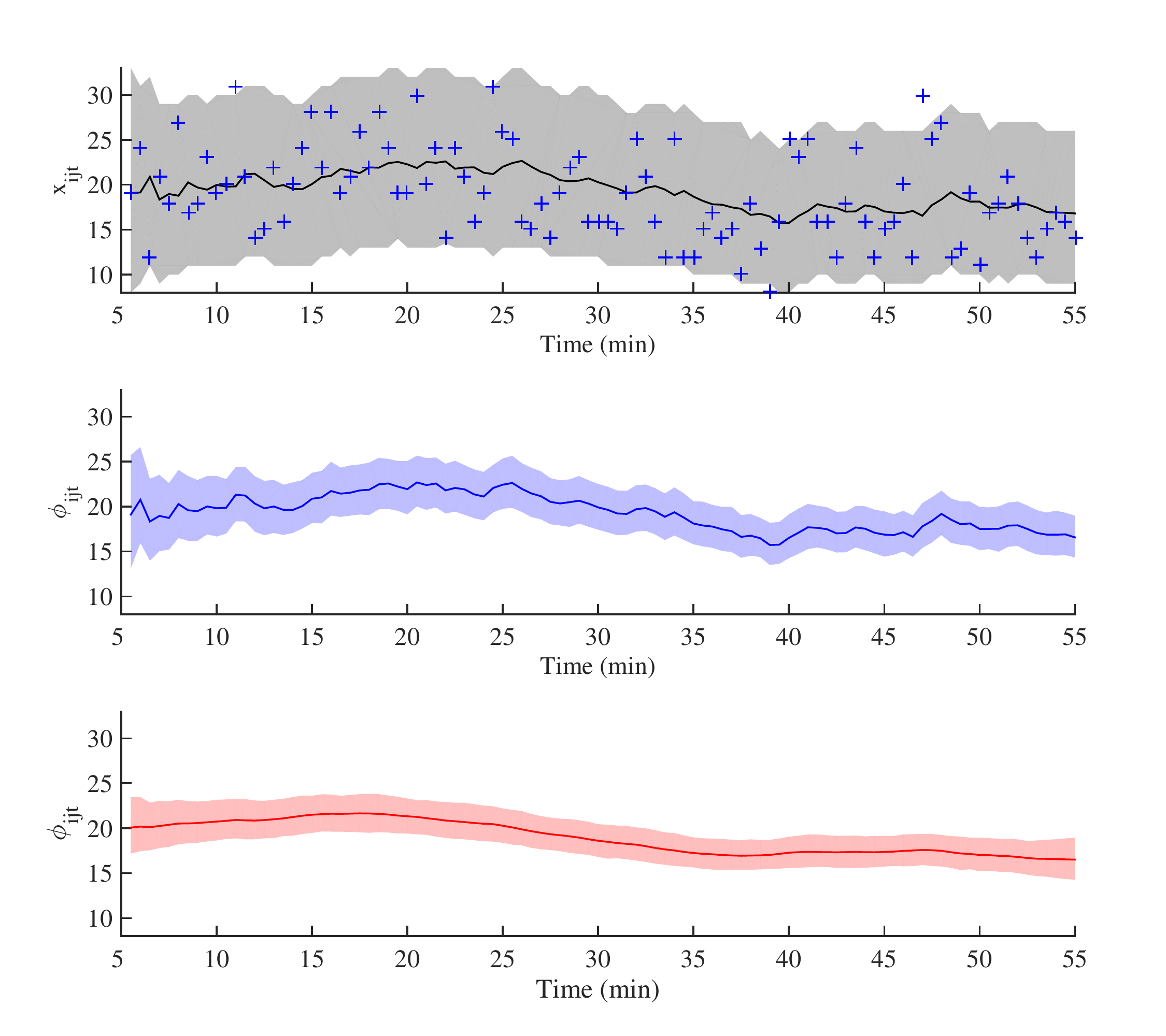} 
\caption{\small BDFM-based inference on inflows to node $j=10$, the Leisure domain. 
{\em Upper:}  data $x_{0,10,t}$ (circles) with one-step ahead forecast means and 95\% intervals.
{\em Center.}  trajectory of mean and 95\% intervals from on-line posteriors $p(\phi_{0,10,t}|x_{0,10,0{:}t})$. 
{\em Lower:}  revised trajectory under full retrospective posterior $p(\phi_{0,10,t}|x_{0,10,0{:}T}).$
}
\label{fig:firsteg}
     \includegraphics[width=3.8in]{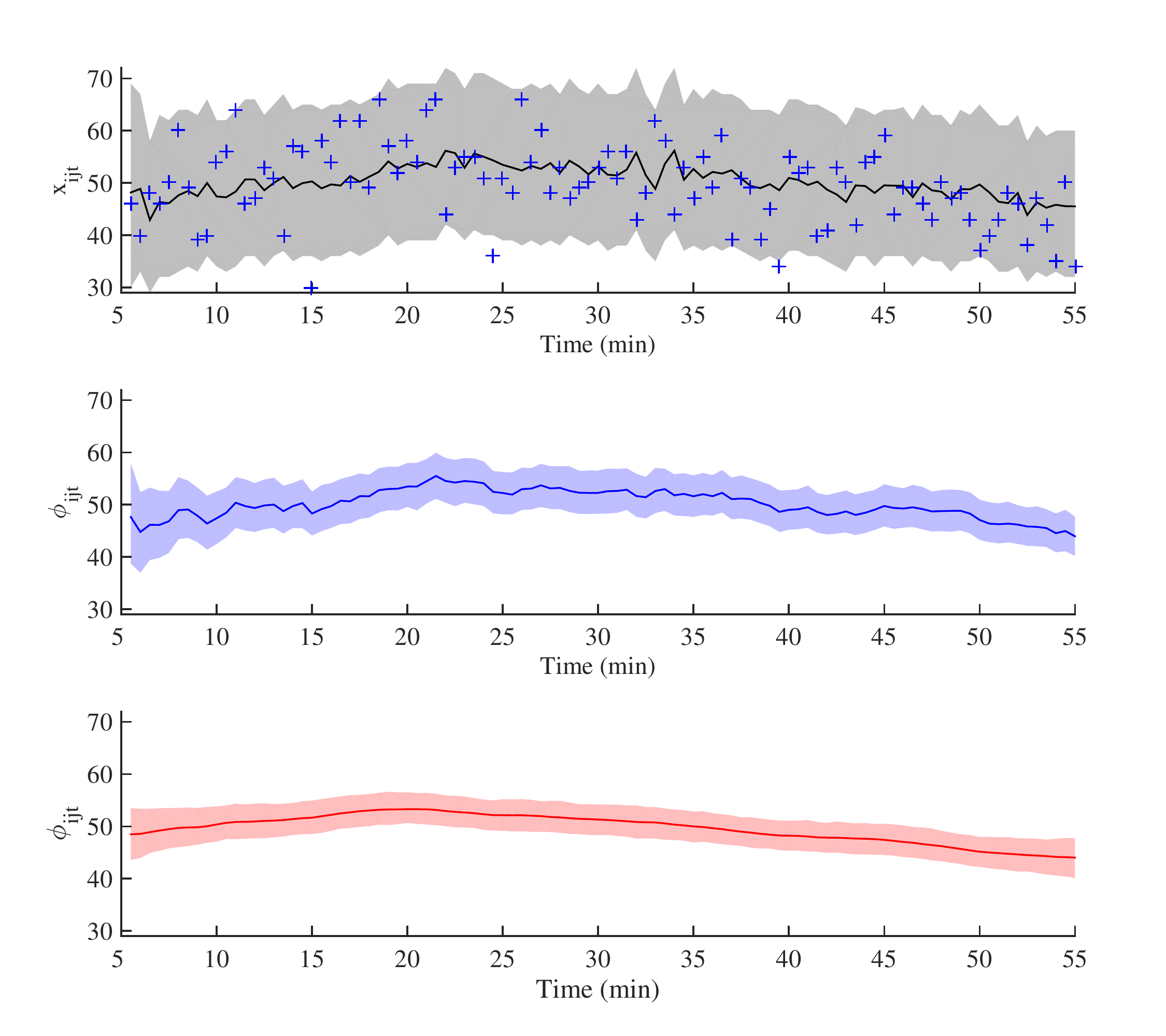} 
\caption{\small BDFM-based inference on transitions from $i=1$ (Homepage) to $j=2$ (Politics). 
{\em Upper:}  data $x_{12t}$ (plus signs) with one-step ahead forecast means and 95\% intervals.
{\em Center.}  trajectory of mean and 95\% intervals from on-line posteriors $p(\phi_{1,2,t}|x_{1,2,0{:}t})$. 
{\em Lower:}  revised trajectory under $p(\phi_{1,2,t}|x_{1,2,0{:}T}).$
}
\label{fig:phihomepol}
\end{figure}

Summary inferences on selected model components are reported
from models with discount factors set at posterior modes.  
Figure~\ref{fig:firsteg} gives one example of learning about inflow 
the Leisure domain. 
This exemplifies sequential learning about the flow
rate together with its retrospectively updated trajectory and a 
visual assessment of one-step ahead forecasting aligned with the data. 
A similar display in Figure~\ref{fig:phihomepol} is an example of flow between two 
network nodes: from Homepage to Politics. 
As with Figure~\ref{fig:firsteg}, we note the 
concordance of incoming data with the one-step predictive intervals as they
are successively revised in the forward analysis, and the enhanced smoothing of 
trajectories in the retrospective analysis. 
\begin{figure}[htbp!]
 \centering 
    \includegraphics[width=5in]{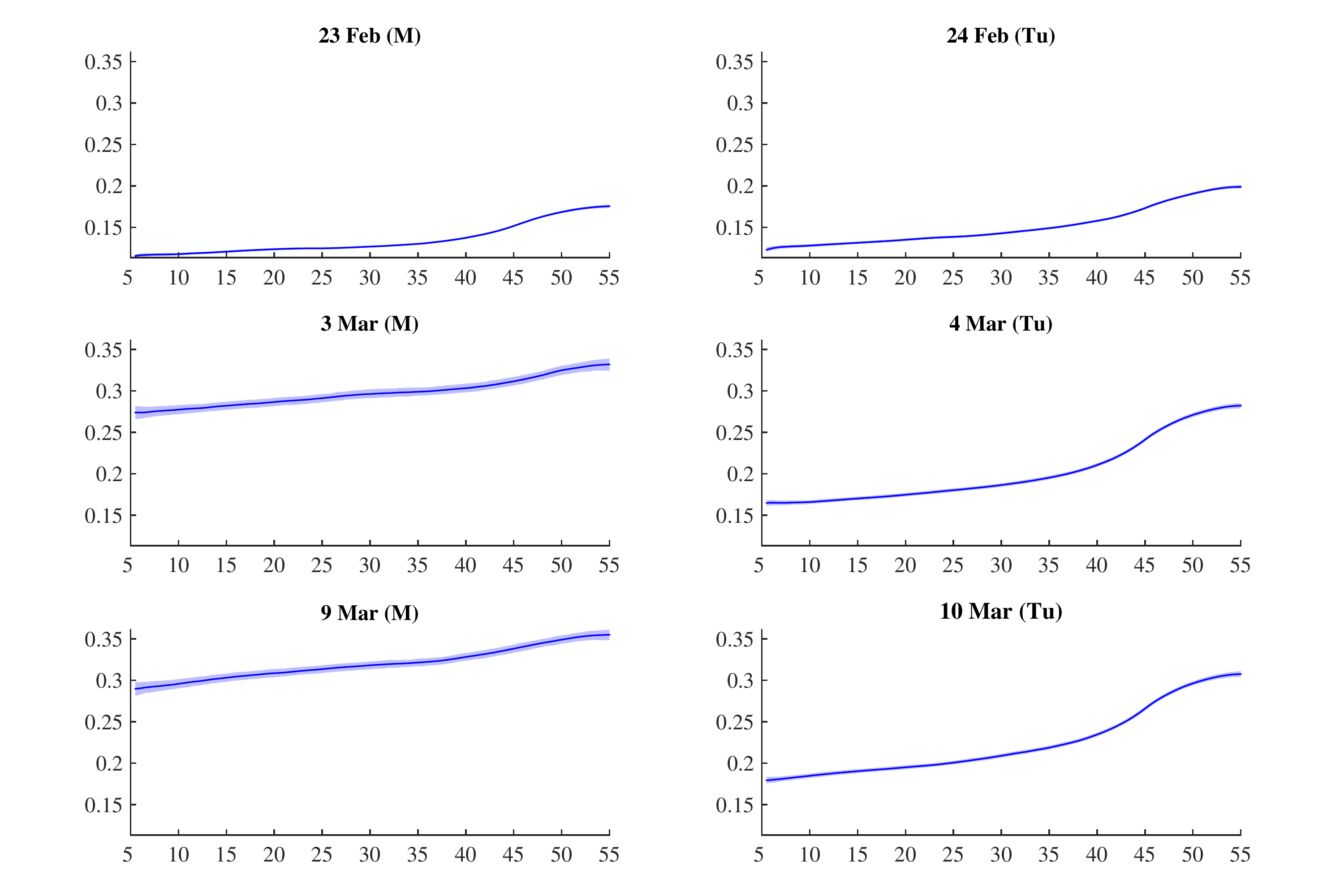} 
  \caption{\small Retrospective posterior means and $95\%$ pointwise intervals of the trajectories of
    transition probability $\theta_{1,0,t}$ (Homepage $\rightarrow$
    External) from analysis on each of the six mornings.}  
 \label{fig:trans2}        
    \centering 
      \includegraphics[width=5in]{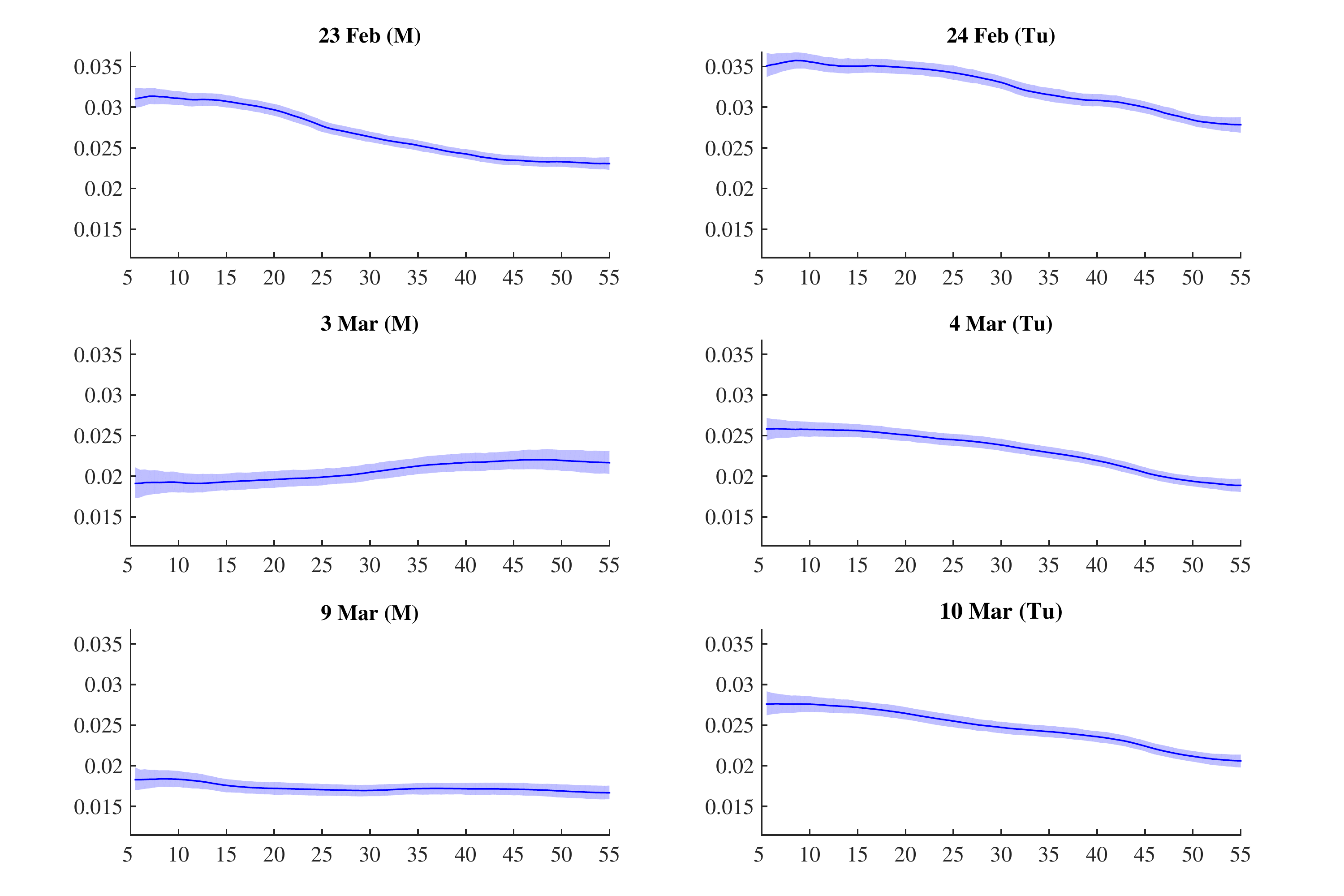} 
  \caption{\small Retrospective  posterior means and $95\%$ pointwise intervals of the trajectories of transition probabilities 
 $\theta_{1,5,t}$ (Homepage $\rightarrow$ Entertainment) for each of the
 six mornings. } 
 \label{fig:trans5}        
\end{figure}

On transition probabilities, it is natural to look at examples
involving the Homepage,  the most popular single domain on Fox News. 
For example, Figure~\ref{fig:trans2} shows that the probability of people leaving 
the Fox News website from Homepage  increases in this 50 minute 
window for each of the six mornings. Note that there are significant day effects; 
e.g., visitors were more likely to leave Fox News on the morning of
March 9th compared to the other mornings.  More detailed insights based on the gravity model
analysis are noted in Section~\ref{sec:DGMdata}. 
Similar figures (not shown) highlight patterns and day-to-day 
differences for other transitions. For example, 
most Homepage visitors stay on Homepage for a while and have a high probability of 
exiting the Fox News site entirely from that page.  
Across all six days, the probability of staying on the Homepage each time interval generally
decreases over the course of the 50 minute morning period.

As an illustration of a more detailed analysis of a very specific flow,
consider Figure~\ref{fig:trans5}.
Among the visitors who leave the Homepage for other Fox News 
domains, Entertainment is generally the most popular destination. 
Across all mornings, we see major differences in trajectories of transition probabilities; in particular
February 23rd and 24th have higher rates than the other days.
It is noteworthy that the Academy Awards ceremony was held on the
night of February 22nd, which may have driven this uptick.
Additional summaries of  inference on trajectories of selected 
$\phi_{ijt}$ and $\theta_{ijt}$  processes appear in the mapping to dynamic gravity models;
see the upper row of frames in Figures~\ref{fig:gmHome_Sci} and \ref{fig:gmHome_Ent}, for examples.

\section{Emulation-Based Mapping to Dynamic Gravity Models \label{sec:DGM}}

\subsection{Dynamic Gravity Models}

A more intricate, multivariate dynamic model  involves node-specific main effects and node-node interaction terms,  representing dependencies in patterns of flows linked to 
inflow/outflow and node-node relationships. 
For each within-network node $i=1{:}I$ and all  $j=0{:}I,$   the model is
\begin{equation}\label{eq:DGM} 
 \phi _{ijt} = \mu _t \ \alpha _{it} \ \beta _{jt} \ \gamma _{ijt}
\end{equation}
with:  (i) a baseline  process $\mu_t$; (ii) a node $i$ main 
effect process $\alpha_{it},$ adjusting the baseline  intensity of flows-- the origin or outflow parameter process for node $i$; 
(iii)  a node $j$ main effect process $\beta_{jt},$ 
representing the additional ``attractiveness" of node $j$ -- 
the destination or inflow parameter process for node $j$; and (iv) an
interaction term $\gamma_{ijt}$, representing the directional
``affinity" of node $i$ for $j$ over time relative to the combined contributions of 
baseline and main effects. 
 
Models of this and more elaborate forms have seen some use in 
transportation studies~\citep[e.g.][]{West1994, Sen:1995} where the 
interaction term is typically structured as a function of physical distance
between nodes; there the ``gravity model" terminology relates to 
the role of small distances in defining large interactions and 
hence ``attraction" of traffic from node $i$ to node $j.$ 
We refer to the $\gamma_{ijt}$ interactions as ``affinities".  
In dissecting the network flow activity, we are most 
interested in questions about which affinities are greater than one 
($j$ attracts flow from $i$ over and above the main effects), or less 
than one ($j$ is relatively unattractive to $i$), or not significantly 
different to one (neutral). 
Critically, affinities are time-varying, and any identified patterns 
of variation over time may be related to 
interpretable events or network changes. 

In a first fully Bayesian approach to gravity models using 
MCMC methods, 
\citet{West1994} developed such models in the static case; i.e., 
with no dynamics in the model parameters, and applied the model
to a large transportation flow network.  
\cite{Cogdon2000} explored a similar approach in studies of patient 
flows to hospitals.  
Analysis via MCMC is computationally very demanding, and the burden 
increases quadratically in $I$, and inherently non-sequentially. 
More recently, \cite{JandarovEtAl2014} studied such models for spread
of infectious diseases, and used Gaussian process 
approximations for approximate inference rather than full MCMC or
other computational methods. 

We share the spirit of this latter work, in using the simply and 
efficiently implemented BDFM as a path to fitting the gravity 
model---now extended to time-varying effect parameter processes. 
However, we do not constrain the affinity parameters $\gamma_{ijt}$ 
as a function of covariates of any kind, simply treating the DGM 
as a dynamic, random effects model.   
This leads to a {\em direct} parameter mapping between the 
BDFM to the DGM; as a result, the trivially generated simulations 
from the full posterior of the BDFM are mapped directly to full 
posterior samples from the DGM, providing immediate access to 
inference on main effect and affinity processes over time.

\subsection{Model Mapping for Bayesian Emulation of DGMs by BDFMs} 
 
Given a set of flow rates $\phi_{ijt}$ for all $i=1{:}I, j=0{:}I,$ at
each time $t=1{:}T,$ the mapping to DGM parameters in \eqno{DGM} 
requires aliasing constraints to match dimensions. 
We adopt the common zero-sum constraint on logged values. 
Define $h_{t}=\log(\mu _t)$, 
$a_{it}=\log(\alpha_{it})$, $b_{jt}=\log(\beta_{jt})$ and 
$g_{ijt}=\log(\gamma_{ijt}).$
Using the $\bigcdot$ notation to denote summation over the range of
identified indices,  constrain via
$a_{\bigcdot t}=b_{\bigcdot t} =0$,  $g_{\bigcdot jt}=g_{i\bigcdot t}=0$ 
for all $i,j,t.$
We then have a bijective map between BDFM and DGM parameters;
given the $\phi_{ijt}$ we can directly compute implied, identified DGM parameters.
The  DGM is saturated-- there are exactly as many parameters in the
DGM as there are observations in the data set. However, the emulating 
BDFM enforces smoothness over time in parameter process trajectories, and this
acts to substantially reduce the effective   model dimension-- one key attribute of the
emulation approach.   Note that this overall strategy inherently adopts the view that 
temporal structure for DGM parameter processes are those induced by the mapping from BDFMs. 
In current form, the evolution of latent rate processes in the latter are random
walks with levels of variation defined by rate-specific discount factor sequences, so the
evolutions for the induced DGM parameters will be more elaborate but still basically
of random walk form.

Define $f_{ijt}=\log(\phi_{ijt})$ 
for each $i=1{:}I, j=0{:}I,$ at each time $t=1{:}T.$  
Then at each time $t,$ compute the following in the order given: 
\begin{itemize} \setlength\itemsep{0pt}
\item the baseline level $\mu_t=\exp(h_t)$ where $h_t =
  f_{\bigcdot\bigcdot t}/I(I+1)$; 
\item for each $i=1{:}I,$ the origin node main effect
$\alpha_{it}=\exp(a_{it})$ where $a_{it} = f_{i\bigcdot t}/(I+1) -
h_t$; 
\item for each $j=0{:}I,$ the destination node main effect
  $\beta_{jt}=\exp(b_{jt})$ where $b_{jt} = f_{\bigcdot jt}/I - h_t$; 
\item for each $i=1{:}I$ and $j=0{:}I,$ the affinity
  $\gamma_{ijt}=\exp(g_{ijt})$ where $g_{it} =f_{ijt} - h_t - a_{it} - b_{jt}.$
\end{itemize}
In our data analysis below, we apply this to all simulated 
$\phi_{ijt}$ from the full posterior analysis under
the BDFM to map to posteriors for the DGM parameter processes. 

A technical problem with this mapping arises in cases of sparse 
flows, i.e., when  multiple $x_{ijt}$ counts are zero or very small 
for multiple node pairs.     
In such cases the posterior for $\phi_{ijt}$ favors very small 
values and the log transforms are large and negative, which unduly 
impacts the resulting overall mean and/or origin or destination means.
While one can imagine model extensions to address this, at a 
practical level it suffices to adjust the mapping as is typically
done in related problems of log-linear models of contingency tables 
with structural zeros \citep[chap. 5]{Bishop:1975}.  
This is implemented by simply restricting the summations in 
identifiability constraints to  node pairs for which  
$x_{ijt}>d$, for some small $d,$ and adjusting divisors to 
count the numbers of terms in each summation. 
For our study, we use $d=3$. 
With this adjustment, very small $\phi _{ijt}$ appropriately 
lead to very small affinities $\gamma _{ijt}$, i.e., 
small rates underlying very sparse flows.

\subsection{Aspects of DGM Analysis of Fox News Data \label{sec:DGMdata}}

\subsubsection{February 23rd 2015, 09{:}00--10{:}00am.}

We first apply the gravity model decomposition to the morning data 
on February 23rd.  Following BDFM analysis as in Section~\ref{sec:BDFMdata},
posterior simulations (5{,}000 Monte Carlo samples) of flow rates are mapped to posterior samples from the 
corresponding dynamic gravity model. 
%

\begin{figure}[p!]
    \centering 
     \includegraphics[width=6.5in]{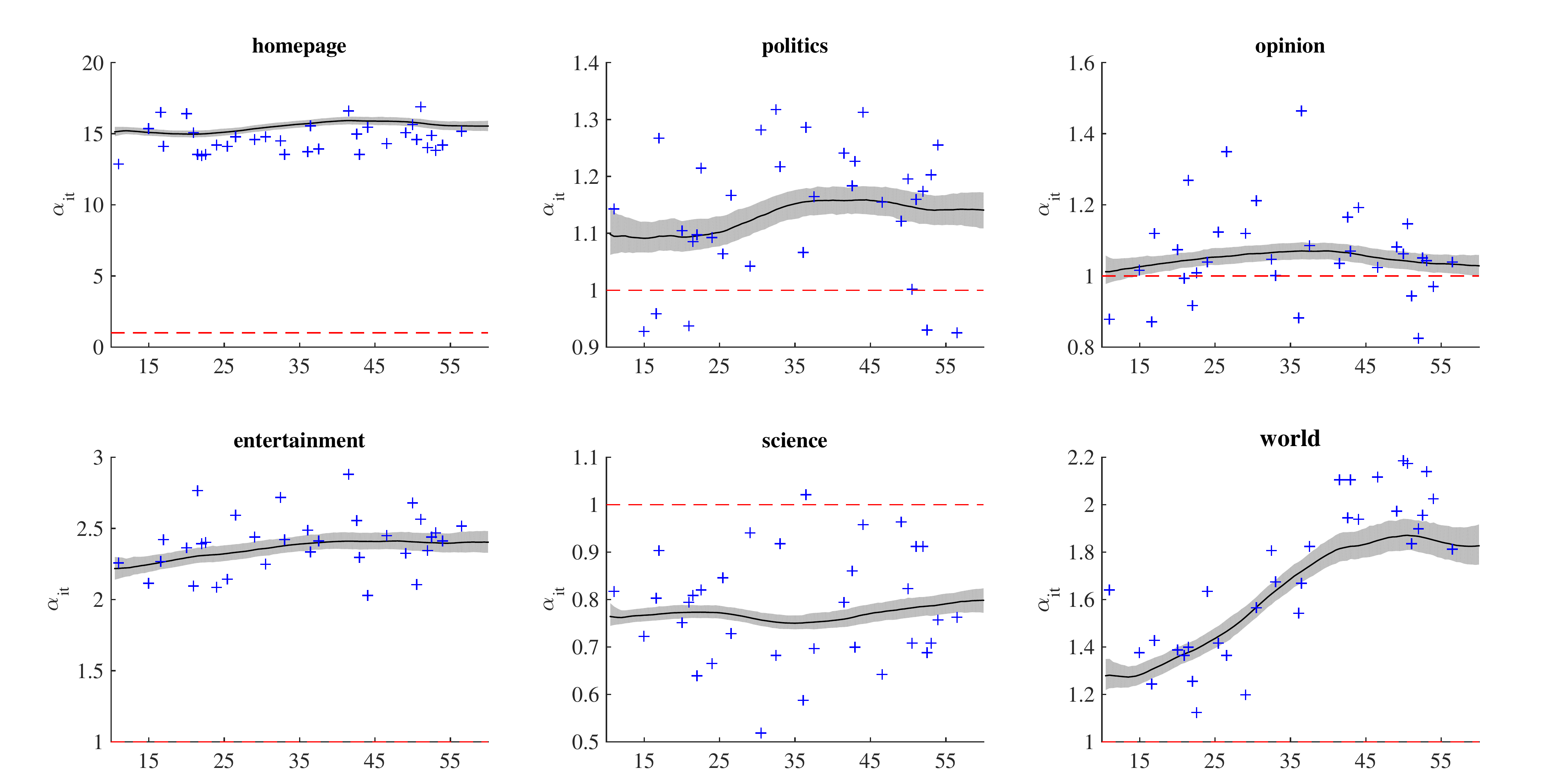} 
  \caption{\small Smoothed trajectories of selected node-specific outflows $\alpha_{i,1{:}T}$.
The + symbols indicate empirical values computed from the raw data (with cases of 
0 occupancy leading to missing values). 
} 
\label{fig:alphat}
    \centering 
     \includegraphics[width=6.5in]{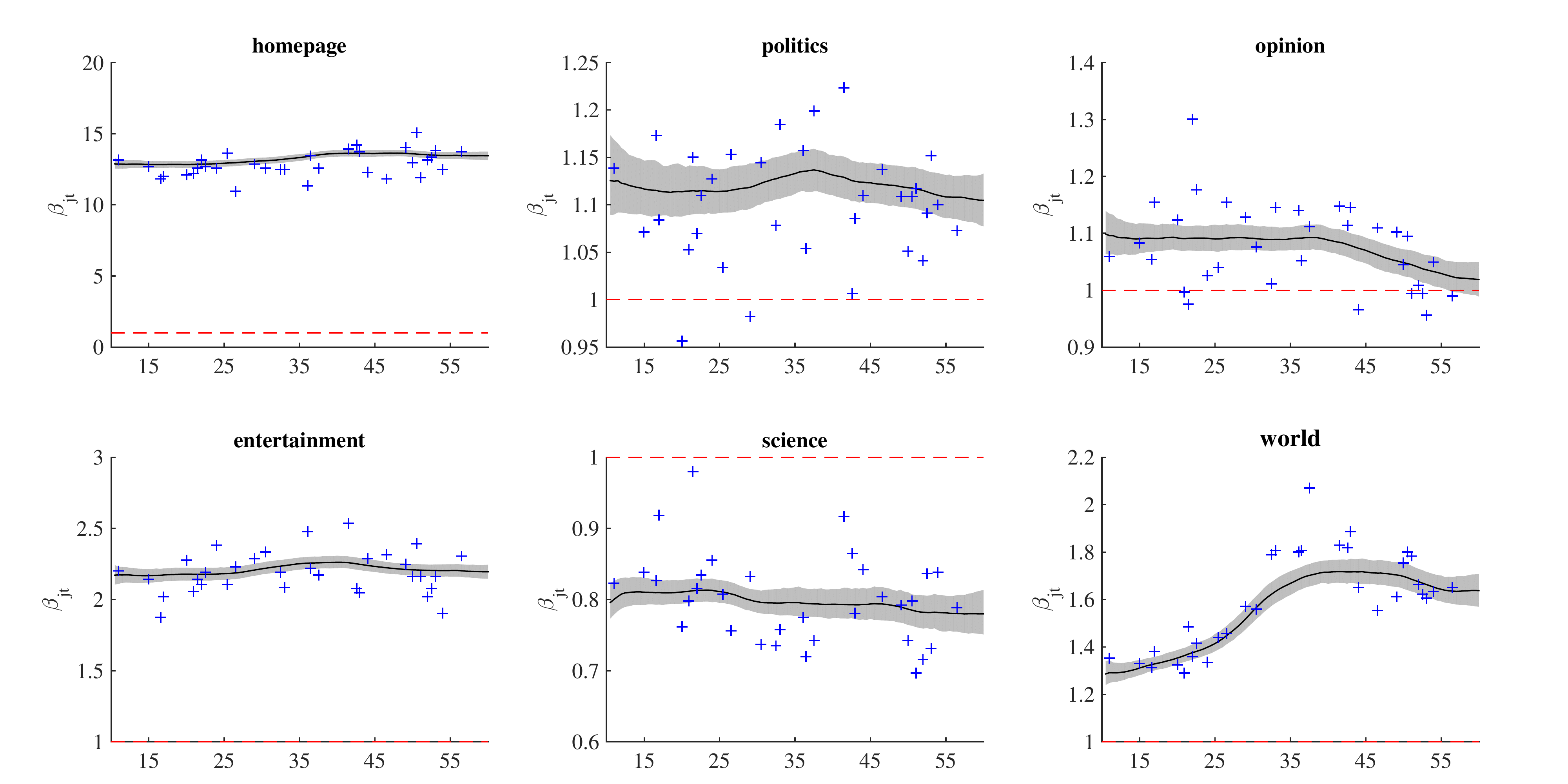} 
  \caption{\small Smoothed trajectories of selected node-specific inflows $\beta_{j,1{:}T}$.} 
\label{fig:betat}
\end{figure}

Inference on outflow (origin) parameters $\alpha_{it}$ 
and inflow (destination) parameters $\beta_{jt}$ for six chosen nodes are shown in
Figures~\ref{fig:alphat}~and~\ref{fig:betat}. 
The posteriors for origin effects show that large-scale domains,
such Homepage, have higher values of $\alpha _{it}$, 
while domains with low or zero flows, such as Science, naturally  have lower values.
Across all domains, subsets  show similar patterns but there are also major
differences apparent. In particular, the posterior analysis shows that several domains, such as
Homepage and Entertainment, are substantially higher than the average as both origin and destination nodes. 
Several nodes, such as Opinion, have above (or below) average destination effects but origin 
effects about the norm.   These distinctions between the two effects show the roles of
$\alpha _{it}$ and $\beta _{jt}$ as 
representing common factors across the origin and destination of
the flows node-by-node. 
They are also naturally related over time in most domains; this captures  the effect of 
the overall scale, or popularity, of some domains such as  Homepage and Entertainment here,
while also showing up in clearly similar patterns over time in less active domains, such as World. 
Further, while some trajectories are relatively stable over time, others show marked changes 
in the node-specific effects over the morning period.  Opinion, for example, has a 
roughly constant and above-average inflow effect for much of the morning,  but it decays toward 
the end of the morning period; World starts off at a level slightly above the norm in both inflow and 
outflow effects,  which both then increase substantially as the morning progresses; Science, in contrast, 
has roughly constant effects across the full time period.

For the affinity effects $\gamma _{ijt}$, we have $(I+1)^2-1$
parameters (one for each pair of nodes except the unobserved External
$\rightarrow$ External flow) at each time $t$. 
The number of effects becomes massive for large $I$.
Even in this example for illustration, $I=22$, the number of 
$\gamma_{ijt}$ for fixed $t$ is 528, so it is impossible to examine 
all the results in this paper. 
For this reason, we pick up a few affinity effects that may interest 
readers in terms of interpretation.
For affinity $\gamma_{ijt}$ with retrospective posterior 
c.d.f $\Phi_{ijt}(\gamma)$, we introduce the {\em Bayesian credible value}
$p_{ijt}=\mathrm{min}\{\Phi_{ijt}(1),1-\Phi_{ijt}(1)\}$
as a simple numerical measure of deviation from the ``neutral" 
value of 1.
This highlights the practical relevance of the affinity effect and 
its changes over time. 

\begin{figure}[t!]
    \centering 
     \includegraphics[width=6.5in]{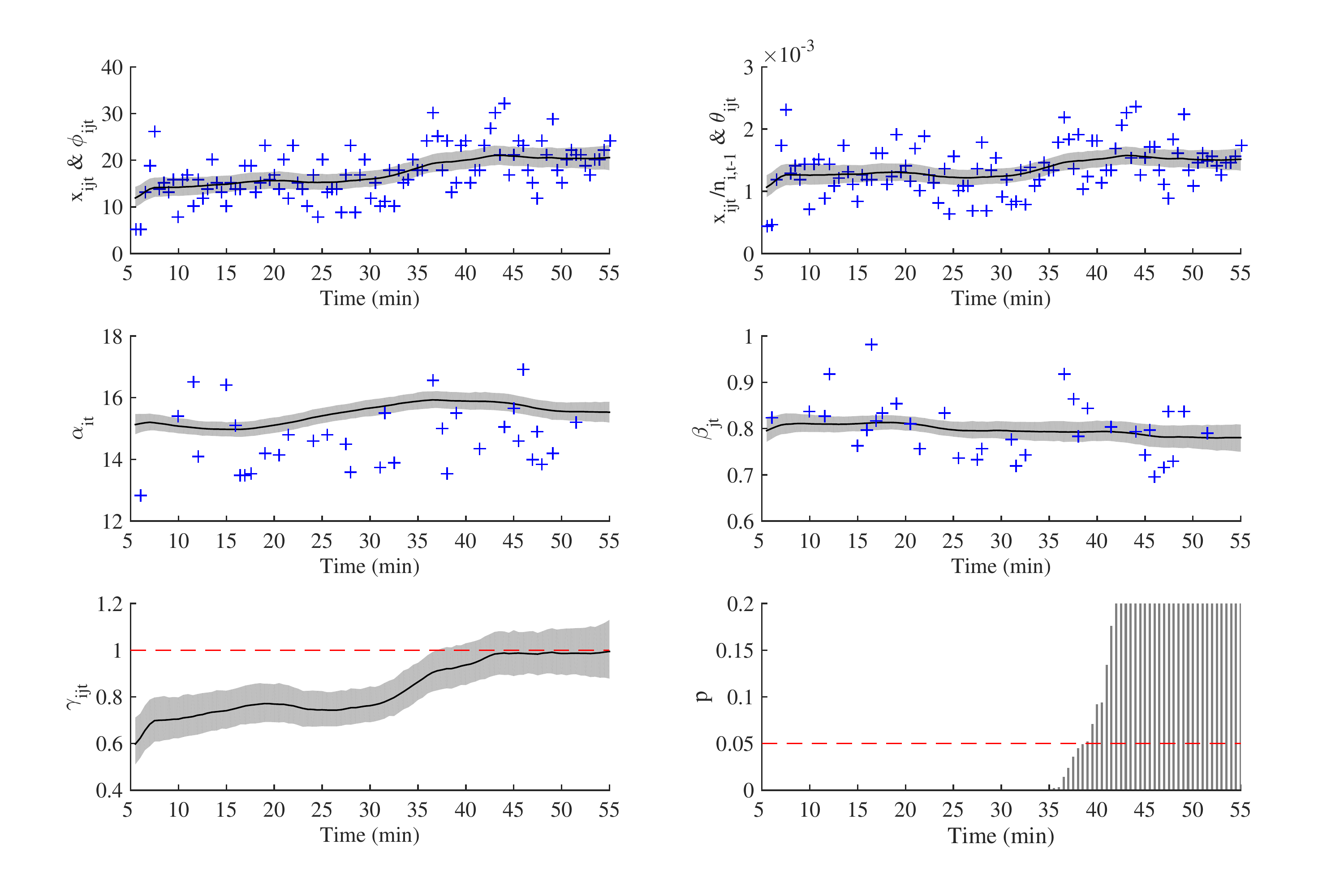} 
\caption{\small   Posterior summaries for DGM parameters for transitions from
node $i$=Homepage $\rightarrow j$=Science. As in Figures~\ref{fig:alphat} and \ref{fig:betat}, 
the + symbols indicate empirical values  (with cases of 
0 occupancy leading to missing values). 
{\em Upper left:}  Posterior trajectory for the latent flow level process $\phi_{ijt}$ with raw counts (crosses). 
{\em Upper right:} Posterior trajectory for the transition probability process $\theta_{ijt}$ with 
raw frequencies (crosses). 
{\em Center left:}  Posterior trajectory for the Homepage origin (outflow) effect process $\alpha_{it}.$ 
{\em Center right:}  Posterior trajectory for the Science destination (inflow) effect process $\beta_{jt}.$ 
{\em Lower left:}  Posterior trajectory for the Homepage:Science affinity/interaction
process $\gamma_{ijt}$.  
{\em Lower right:} Corresponding trajectories of Bayesian credible values 
assessing support for $\gamma_{ijt}$ near 1. 
}
\label{fig:gmHome_Sci}
\end{figure}

Traffic from Homepage to other domains are central to 
understanding normal patterns of variation as  Homepage is usually the landing
page for visitors. Where users tend to go next, and how the flow patterns 
begin to evolve from Homepage generally, is one key interest from the 
advertisement and marketing viewpoints.
Figure~\ref{fig:gmHome_Sci} displays some relevant summaries for flows from Homepage to Science.
First, overall counts and also relative frequencies of transitions tend to increase over this
morning period. The BDFM appropriately tracks these slowly evolving trends
(while not, of course, predicting them). Second,   origin/outflow and destination/inflow 
parameter processes are relatively constant over time,  although the former exhibits 
a slight increase in the later morning period.  Of more interest-- and highlighting the 
flexibility and incisiveness of the BDFM$\rightarrow$DGM emulation map-- the affinity 
process is clearly time-varying. Initially at reduced levels-- Fox News visitors tend to be much less 
likely to go to Science from Homepage during the first half or more of the morning period-- 
this trends upwards to be basically negligible in impact  after about 35-40minutes. Note that, 
while the overall outflow and inflow processes are roughly constant over time for this
par of nodes, the raw data indicate continued growth in traffic towards the end of the time
period, and thus the model responds by inferring the upward drift in the interaction/affinity process. 

The pattern over time of the affinity effect also relates to  \lq\lq dynamic sparsity." 
While we do not have models 
that are explicitly exploring sparsity in main or interaction effects,  the emulation approach has
enabled the identification of an interaction/affinity process  that is relevant for some
period of time but then, practically, irrelevant for others. In contrast to our easy and 
scalable methodology, other more formal Bayesian approaches to dynamic 
sparsity modeling~\citep[e.g.][]{Nakajima2010,Nakajima2011a,Nakajima2014,Zhou2012}
are difficult or impossible to reliably implement in a sequential context, being reliant on
intense MCMC methods for batch data processing.     We do note, however, that we are
not formally testing consistency of posteriors for affinities against the value of 1, but
simply exploring the trajectories to generate insights.  More formal assessment is available,
if desired, by considering differences in affinities over time from the full posterior sample.

\begin{figure}[t!]
    \centering 
     \includegraphics[width=6.5in]{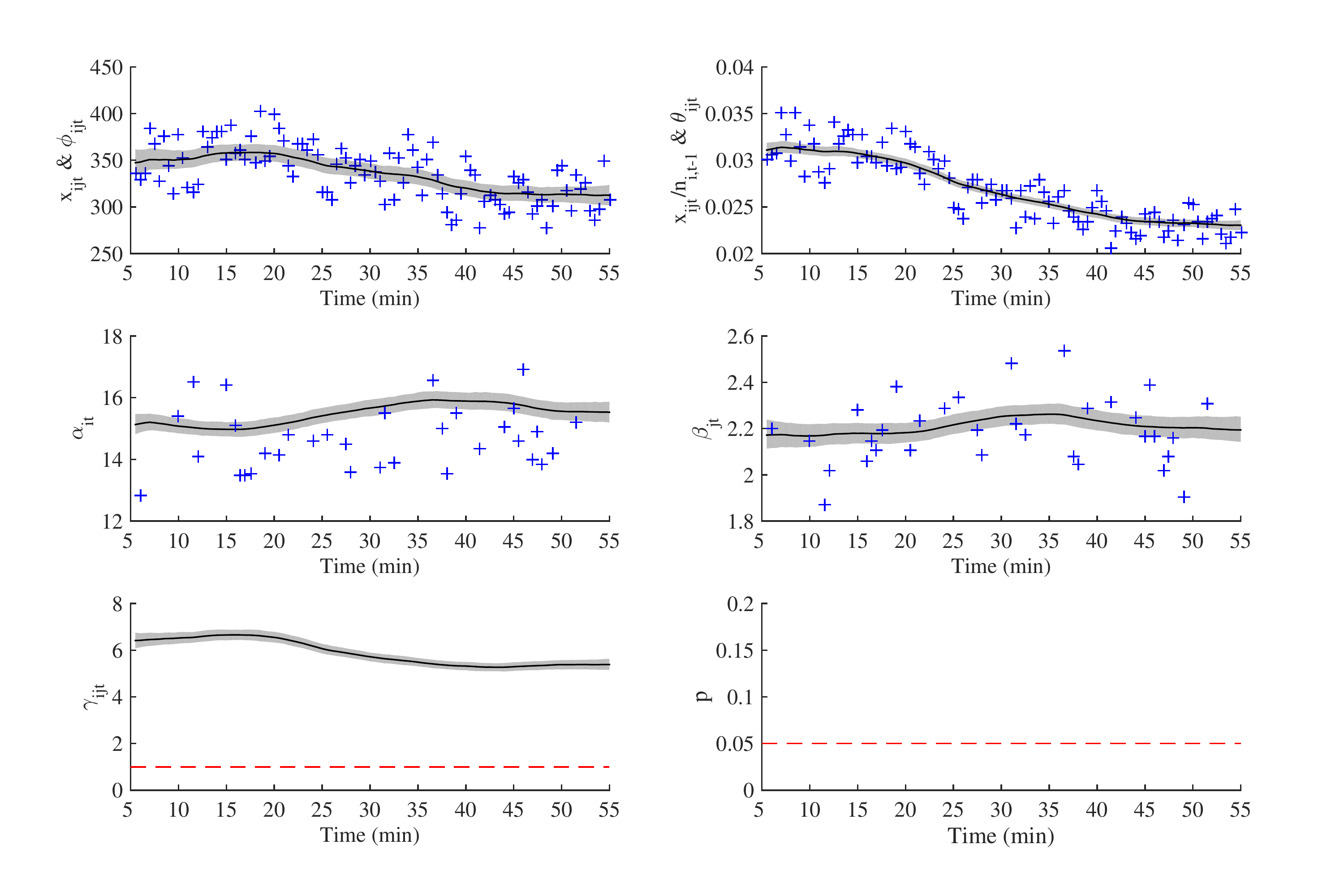}  
  \caption{\small  Posterior summaries for DGM parameters for transitions from
node $i$=Homepage $\rightarrow j$=Entertainment, with details as in Figure~\ref{fig:gmHome_Sci}.
} 
\label{fig:gmHome_Ent}
\end{figure}

A second example, chosen to represent node pairs with high inflow and outflow levels, 
concerns transitions from Homepage to Entertainment;  see Figure~\ref{fig:gmHome_Ent}.
Here again we see that the trivially implemented emulation approach is able to identify a high level of
stability over time in the main effects, while indicating subtle changes to reduced levels of 
affinity in the latter part of the morning period. Considering the downward trending 
patterns in raw data/relative frequencies
of flows from Homepage to Entertainment in the latter period, it would not otherwise be 
easy to isolate these patterns as idiosyncratic to the node pair.  Inferences reflected here
on the trajectory of the affinity process clearly show significantly reduced levels later on,  
with $\gamma_{ijt}$ falling from around 6.5 to around 5.5; relative to the 
network-wide structure, high affinity is maintained
throughout at a practical level, but at a reduced level later on for this node pair.  

\begin{figure}[htp!]
    \centering 
\begin{tabular}{@{}cc@{}}
	\qquad\qquad Outflow effects $\alpha_{it}$ &\qquad\qquad  Inflow effects $\beta_{jt}$ \\
	\includegraphics[width=2.65in]{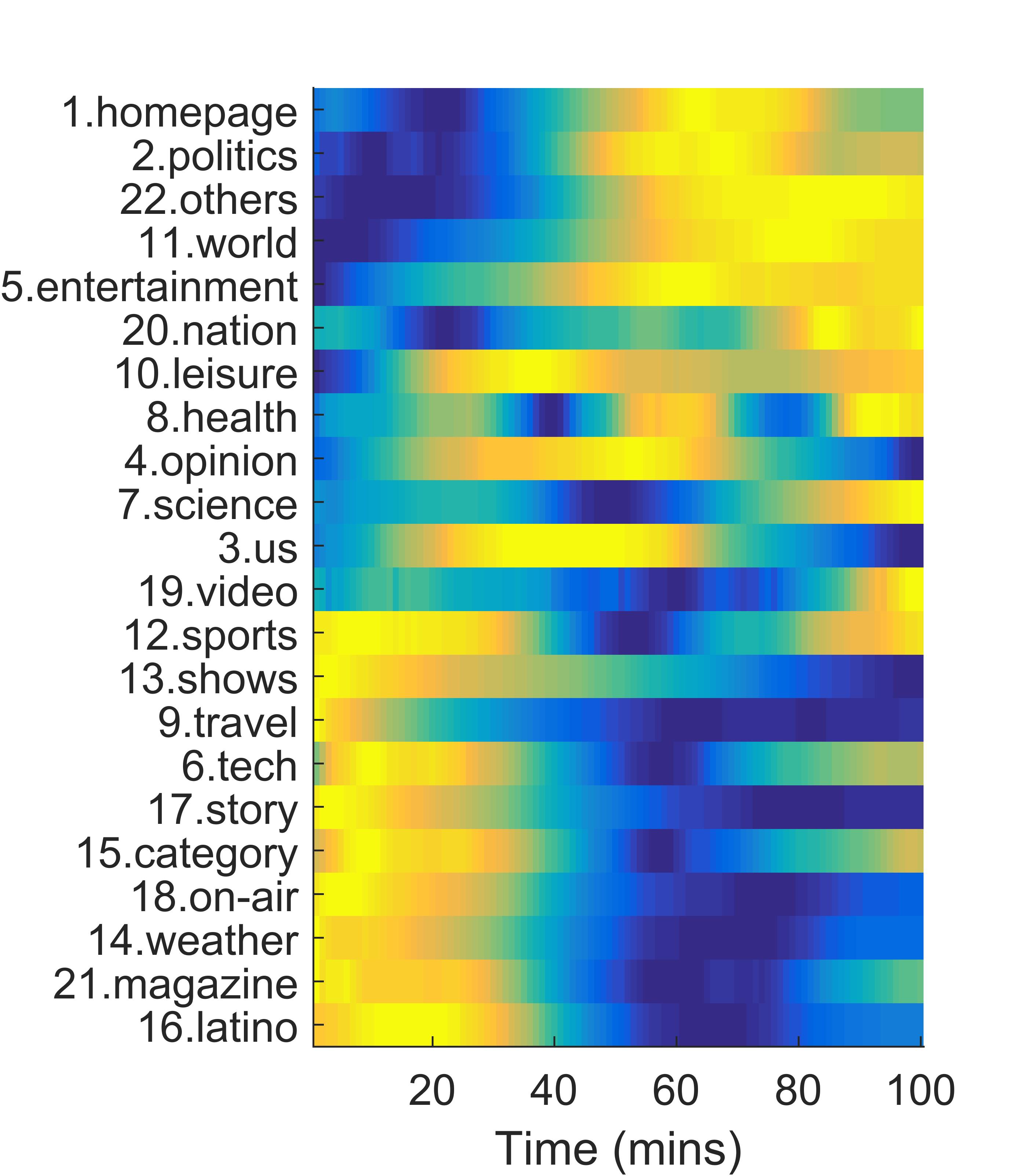}  &
	\includegraphics[width=2.65in]{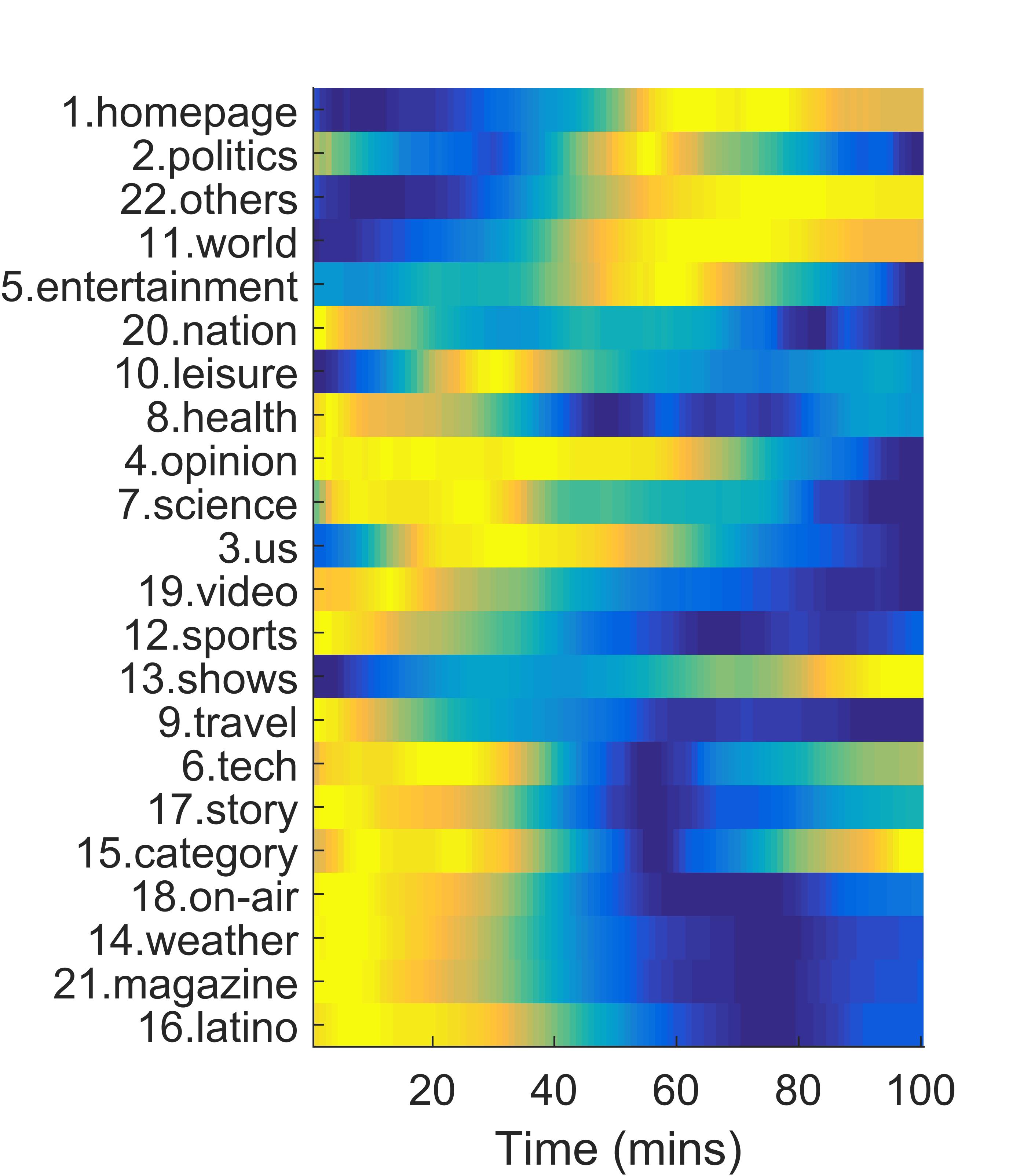}  \\
	\multicolumn{2}{c}{\qquad\qquad Homepage affinities $\gamma_{1jt}$}\\
	\multicolumn{2}{c}{\includegraphics[width=2.65in]{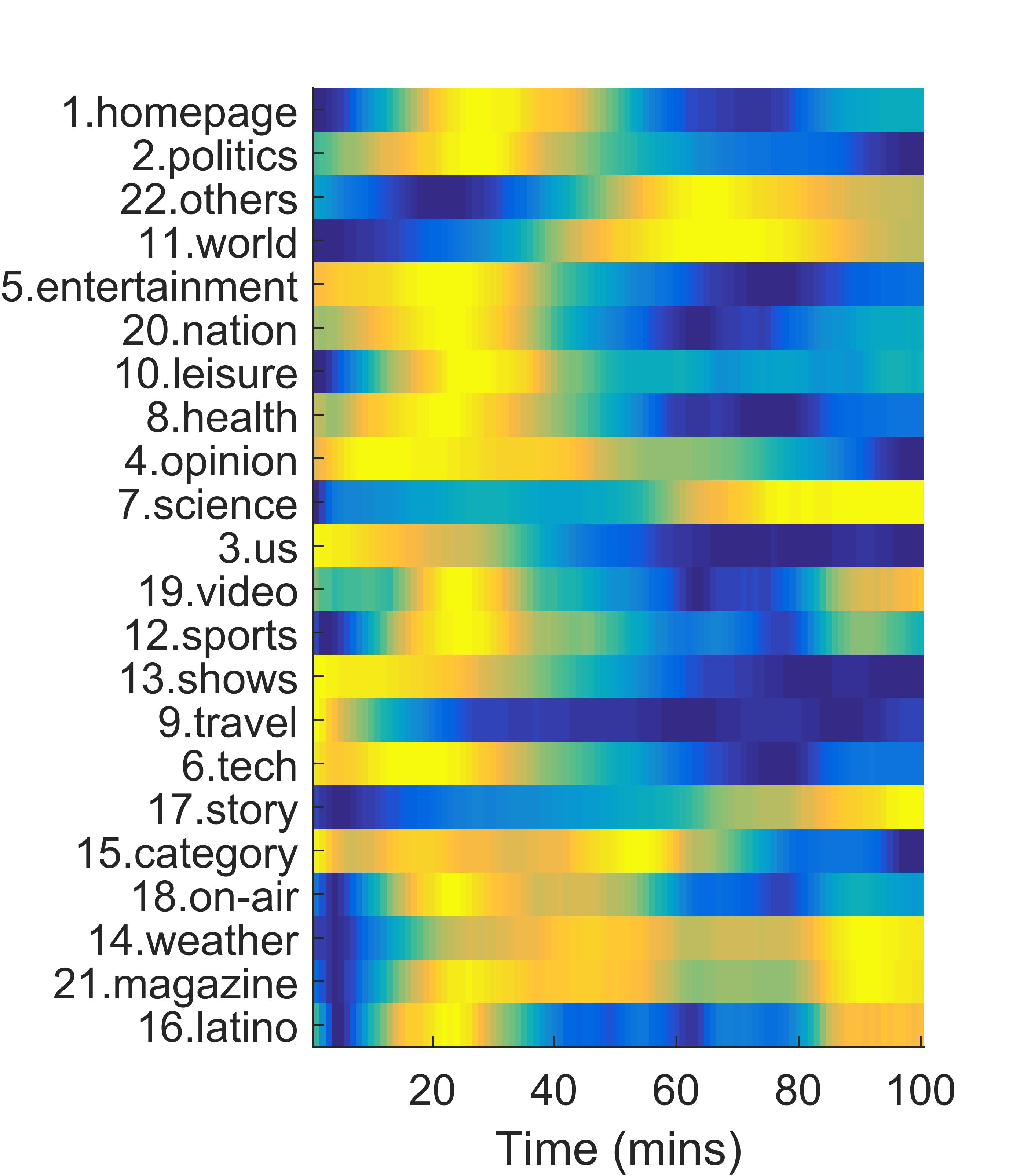} }
\end{tabular}
  \caption{\small  Heat-map images of  
	standardized posterior means of DGM parameters for all nodes across time in the
 	February 23rd am period. 
	Means are max-min standardized separately for each of the $\alpha_{it}$, 
	$\beta_{jt}$ and $\gamma_{1jt}$ (1=Homepage). Shading runs from 0 (dark grey/dark blue in on-line 
	version) to 1 (white/yellow in on-line version) on these standardized scales.  The nodes
 	are ordered based on correlation over time of the main outflow effects $\alpha_{it}$ with
	that of Homepage; this is an arbitrary ordering chosen simply for visual presentation. 
} 
\label{fig:GMimagesFeb23am}
\end{figure}

Some heat-maps in Figure~\ref{fig:GMimagesFeb23am} show aspects of relationships
in some estimated DGM parameters across nodes and across time.  These show patterns in the values of the posterior means of $\alpha_{it}, \beta_{jt}, \gamma_{1jt}$ over time; 
this includes all main effects and the directional affinities/interactions of all network nodes for flows from 
domain $1$, the Homepage.    Simply for \lq\lq nice" visual display,  the nodes are ordered in terms
of correlation over time with the estimated Homepage outflow effect $\alpha_{1t},$ in all three
images. The values shown are standardized within each image so that the min/max across time
are 0/1.  Note common patterns that
reflect interdependencies in dynamics across subsets of network nodes.  The $\alpha_{it}$ 
image reflects natural evolution in the morning period of traffic from network nodes, showing the
increasing rates of transitions from some of the more popular, 
core domains (Homepage, Politics, World, Entertainment  and others) in later morning. 
The $\beta_{jt}$ 
image shows consonant patterns in a subset of these core domains-- in that their 
attractiveness increases in later morning-- but with some clearly different cases. For the 
Homepage affinity processes $\gamma_{1jt},$ there are quite a few domains that see
increased incremental traffic rates in the first half, or so, of the morning period, which then
drop off to low levels later on. 

\subsubsection{Comparison Across Days.}

The study covers 
morning (09{:}00--10{:}00am) and afternoon (01{:}00--02{:}00pm) periods 
on each of 6 days, as already discussed and explored in Section~\ref{sec:BDFMdata}.   Moving to the DGM, 
we now explore additional features concerning  time-of-day effects as well 
as day-to-day variation.  This is based on running the 
coupled BDFM-DGM analysis separately on each time period/day.

\begin{figure}[t!]
    \centering 
     \includegraphics[width=5.5in]{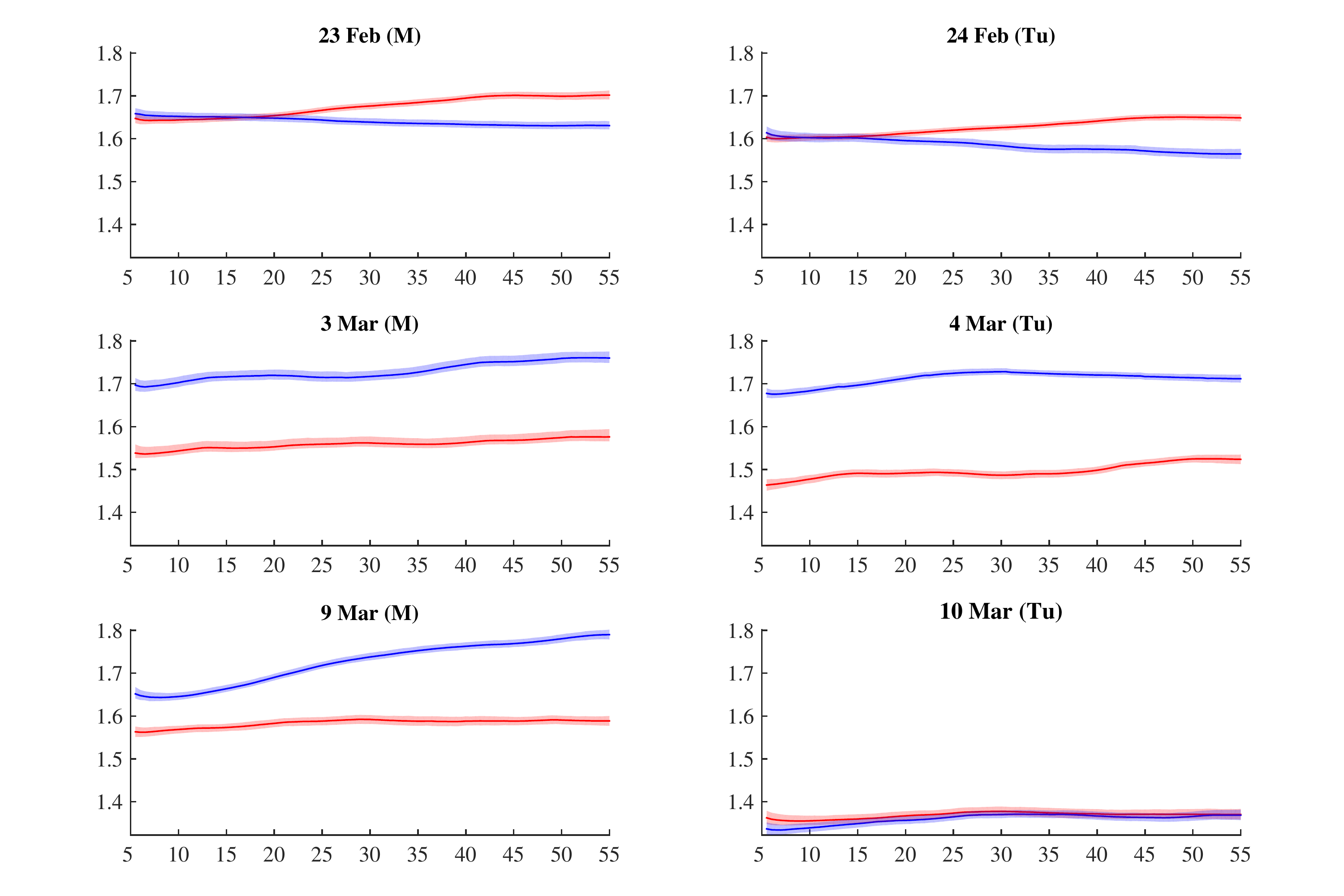} 
  \caption{\small DGM-based inference on trajectories of baseline flow levels  $\mu_t$
across days with 95\%  intervals.  Dark/full lines (red in the on-line paper)
are for morning periods, lighter/dashed lines are for afternoons. 
} 
\label{fig:mdall}
\end{figure}

Figure~\ref{fig:mdall} shows the DGM trajectories for the 
retrospective baseline parameter process $\mu_{1{:}T}$ for each of the 
12 fifty-minute intervals. Trajectories are similar across days but for 
notable differences on February 24th and March 3rd.
On February 24th, the afternoon flow is significantly lower than
the morning flow, while the morning flow that day is much larger than
across other days. 
One plausible reason is increased morning traffic in response to discussions 
following the Academy Awards ceremony, with a resulting lull in the afternoon traffic.
The reverse happens on March 3 where, although the morning traffic seems typical, 
the afternoon traffic is unusually high.
This was the day on which Fox News posted an article concerning
Hillary Clinton's use of her personal email account for all 
correspondence during her tenure as Secretary of State.  
It is plausible that this led to larger than usual
afternoon traffic flows as the controversy unfolded. 

An advantage of the DGM representation is that it allows 
investigation of such speculative explanations.
For example, examination of destination effects (not shown)
$\beta_{5,1{:}T}$ confirm that the Entertainment node was unusually popular
on February 24th am, and that the Politics and Opinion
nodes were unusually popular on March 3 pm, compared
to similar flows on other days.

\section{Sequential Monitoring of BDFMs and Emulated DGMs\label{sec:monitor}}

\subsection{General Comments} 
In routine use of dynamic models in sequential monitoring of flows, one key interest 
is that of being sensitive to data patterns that seem outside the norm, i.e., anomalous, 
and may reflect events and changes requiring investigation and intervention. We address 
this here with methodology based on the concepts and theory underlying 
Bayesian model monitoring in conditionally normal DLMs.  While this theory of 
sequential Bayesian model assessment is well-established, it does not seem to have 
been adapted to apply to dynamic models of counts; our contributions in this paper
include this extension and required customization of the approach.

\subsection{Bayesian Model Monitoring and Automated Intervention} 

Revert to the generic context of a Poisson-gamma model, as in Section\ref{sec:gammabetadm} and
Appendix~\ref{app:betagammadm}, with time $t$ count observation 
$x_t$ and underlying state $\phi_t$. 
The general strategy here applies to all cases: inflows to any node $j$ when $x_t =x_{0jt},$  as
well as transitions from a node $i$ to $j$ when $x_t = x_{ijt}.$ 

The sequential Bayesian testing approach in DLMs~\citep[][Chapter 11]{WestHarrison1997}  
is  extended here to apply to the Poisson-gamma dynamic model. 
Regard the model as the \lq\lq standard model" at time $t,$ relabeling the one-step 
predictive density at each time $t$ as $p_0(x_t|\delta_t,\delta_{1{:}t-1},x_{0{:}t-1});$   the suffix 0 
indicates the standard model, and we now explicitly recognize the dependence on the 
discount factors defining levels of stochastic change in the underlying state process $\phi_t$.  
The two components of monitoring and
adaptation are as follows. 

\medskip\noindent{\em A. Alternative model predictions:} A purely synthetic {\em alternative model} at time $t$ that requires only the specification of 
the alternative predictive p.d.f. $p_1(x_t|\delta_t',\delta_{1{:}t-1},x_{0{:}t-1})$, differing from the
standard only in the current discount factor $\delta_t' < \delta_t.$   This implies that $p_1(\cdot)$ 
is more diffuse that $p_0(\cdot)$ but similarly located. In our Poisson-gamma case, 
 the implied generalized 
negative binomial p.d.f.s $p_0(\cdot)$ and $p_1(\cdot)$ have precisely the same mean but the
latter has a larger variance and gives more support to both smaller and larger values of $x_t$.

\medskip\noindent{\em B. Bayes factor comparisons:} Evaluation of Bayes factors comparing the standard model predictive p.d.f.s with the alternative define the monitoring tests. 
These marginal likelihood ratios are computed based on both the time $t$ observation and 
recent consecutive observations to assess and 
compare consistency of this local data with predictions from 
the standard model relative to the more diffuse synthetic alternative.  Support for the 
standard model is regarded as a \lq\lq business as usual" signal.  A signal of support for the
alternative addresses the potential for: 
(i) the single observation to be discrepant, a possible outlier; (ii) a relatively abrupt change in the $\phi_t$ process at time $t$, beyond that predicted by the model with current discount rate $\delta_t;$ 
and (iii) change in the $\phi_t$ process at higher levels than the norm,  but that are not so abrupt and
may have been developing at subtler levels over a few recent time points. 
 
Define the following:  
\begin{itemize} \setlength\itemsep{0pt}
\item The time $t$  Bayes factor $H_t = p_0(x_t|\delta_t,\delta_{1{:}t-1},x_{0{:}t-1})/
			p_1(x_t|\delta_t',\delta_{1{:}t-1},x_{0{:}t-1}),$	 assessing the current observation alone.  
\item The {\em lag-$h$ local  Bayes factor} 
		    $H_t(h) = \prod_{r=t-h+1{:}t} p_0(x_r|\delta_r,\delta_{1{:}r-1},x_{0{:}r-1})/
			p_1(x_r|\delta_r',\delta_{1{:}r-1}',x_{0{:}t-1}).$
		based on the most recent $h\in \{1{:}t\}$ observations, including $x_t.$ 
\item The {\em local cumulative  Bayes factor}  $L_t = \min_{h\in \{1{:}t\}} H_t(h)$
and corresponding {\em run-length} $l_t$ such that $L_t = H_t(l_t).$   
\end{itemize} 
Bayesian testing theory~(\citealp{West1986a} and 
	chapter 11 of~\citealp{WestHarrison1997}) shows that  the local test measures $L_t, l_t$ are 
trivially updated as time evolves. At time $t$, the updated pair is 
$$
[ L_t,\, l_t ] =
\begin{cases}
	[ H_t, \, 1 ], 	&  \textrm{if } L_{t-1}\ge 1, \\
	[ H_t L_{t-1}, \, 1+l_{t-1} ], &  \textrm{if } L_{t-1}< 1. \\
\end{cases}
$$
Past consistency with the standard model $(L_{t-1}\ge 1)$ means that the
entire focus at time $t$ is on the single observation $x_t.$   If, however,  recent evidence 
weighs against the standard model $(L_{t-1}<1)$,  then 
that evidence continues to accumulate based on the new observation.  The pair $[L_t,l_t]$ 
define a tracking signal  that can be used to formally intervene by rejecting potential outliers 
and adopting the smaller discount factor $\delta_t'$ at such times as well as when $L_t$ and/or
$l_t$ suggest cumulating changes in the $\phi_t$ process beyond the norm.  This operates as follows. 
 
Specify a Bayes factor threshold $\tau$ (e.g. $\tau=0.1$) 
and run-length threshold  $r$ (e.g. $r=4).$   
When standing at time $t$, compute single-period Bayes factor $H_t$. Then: 
	\begin{itemize} \setlength\itemsep{0pt}
		\item If $H_t\le \tau$, reject $x_t$ as potentially outlying. \\
			 $\diamond$ Intervene to apply reduced discount factor  $\delta_t'$ at time $t\to t+1$ in case of changes.
		\item If $H_t>\tau$, then proceed to update $[L_t,l_t]$  to continue monitoring in case of potential changes.
		\begin{itemize} \setlength\itemsep{0pt}
		\item  If $L_t\le \tau$ {\em or}  $l_t\ge r,$ \\
				 $\diamond$ Apply reduced discount factor $\delta_t'$ 
					to allow for adaptation to potential changes; \\
				 $\diamond$ Update using time $t$ data as usual but with this increased prior uncertainty;	\\
						  $\diamond$ Reset monitor to $L_t=1$ and $l_t=1.$ 	
			\item  If $L_t > \tau$ {\em and} $l_t<r,$ \\
					 $\diamond$  Proceed as usual with prior-posterior and monitor updates. 				
		\end{itemize} 
\item
Forecast ahead as desired, then proceed to time $t+1.$	
	\end{itemize}
%
This process is displayed in schematic form in Figure~\ref{fig:flowchart} of Appendix~\ref{app:monitoring}, 
this modified from~\citealp[][Chapter 11]{WestHarrison1997}, which also discusses choices 
of thresholds $(\tau,r)$. We follow the recommendations there for these choices.  As discussed in 
Sections~\ref{sec:inflows} and \ref{sec:transits}, 
the discount factors in the standard models are based on 
 $\delta_t = d+ (1-d)\exp(-k r_{t-1})$ where, in the generic notation here, $r_{t-1}$ is the
shape parameter of the time $t-1$ posterior gamma distribution for $\phi_{t-1}$ and $d$ a 
baseline discount rate.  We therefore select the alternative discount factor $\delta_t'$ 
for the intervention analysis as $\delta_t' = d'+ (1-d')\exp(-k r_{t-1})$ where for some smaller
baseline $d'<d;$ the studies of Fox News network data now mentioned are based on $d=0.1$
whereas the standard models are based on values of $d$ running between 0.9 and 0.99
across the sets of inflow and transition flow models.

\subsection{Fox News Network Example}

\begin{figure}[hp!]
    \centering 
     \includegraphics[width=5.0in]{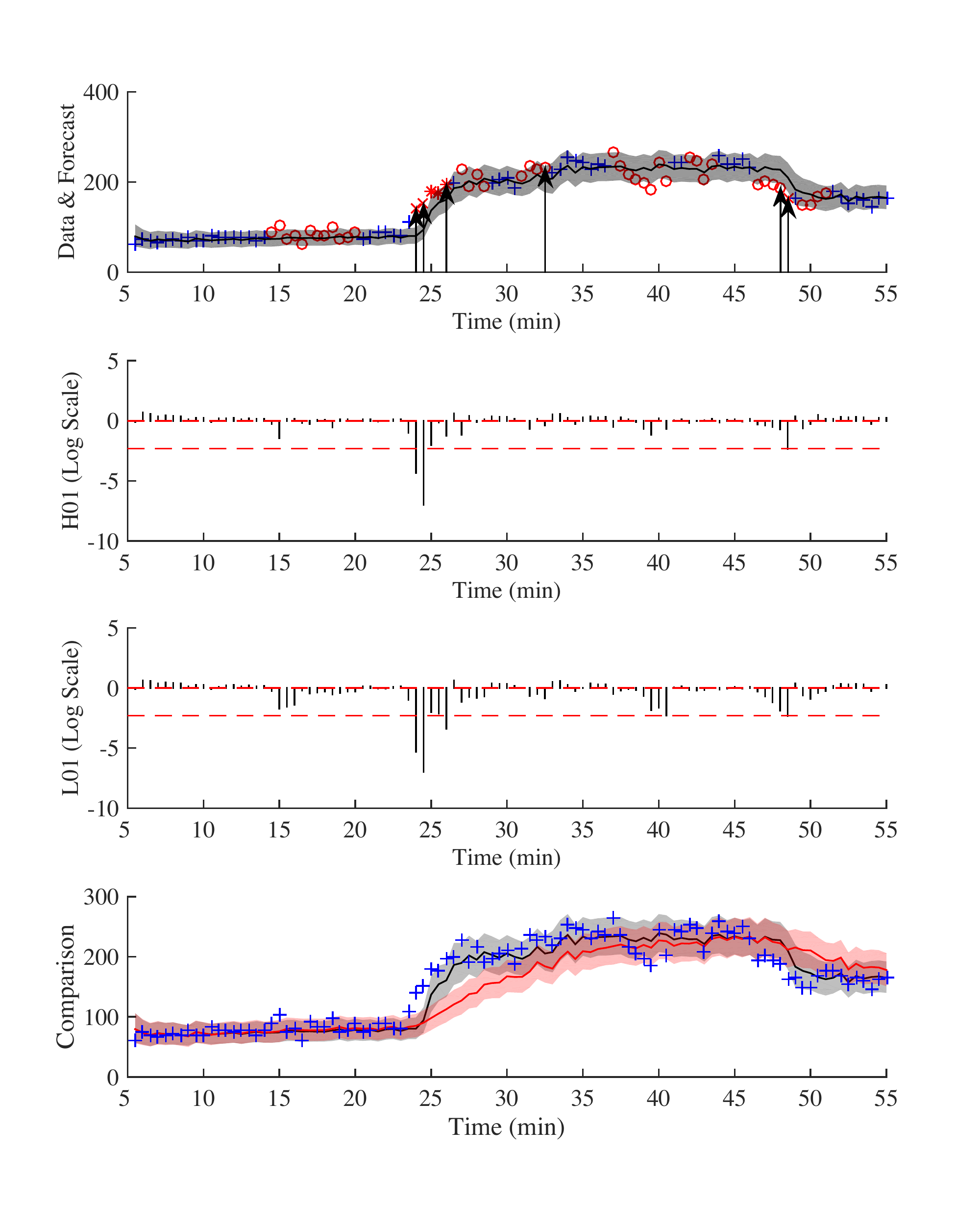}
 \caption{\small  Summaries from analysis of transitions from
node $i$=Homepage $\rightarrow j$=World with Bayesian model monitoring and 
 discount-based intervention. Data is from the February 23rd am period.  
{\em Upper:}  Symbol + indicates observations judged consistent with the 
	standard model; x indicates cases identified as potential outliers by low $H_t$;
	\red{*} indicates those flagged as potential change points via low $L_t$;
	  	\red{o} indicates cases with $l_t>1$.  
	 The vertical arrows indicate times of automatic 
	 intervention. The full line and shaded region represent one-step forecast means and 95\% intervals. 
{\em Center:}  Tracks of $\log(H_t)$ (above center) and $\log(L_t)$ (below center) over time. 
{\em Lower:}  Data (+) with one-step forecast means and 95\% intervals from the 
	standard BDFM analysis in light gray (red in on-line version)
	compared to the analysis with monitoring and intervention in black/dark gray (gray in on-line version). 
} 
\label{fig:monitorHome_World}
\end{figure}

\begin{figure}[hp!]
    \centering 
     \includegraphics[width=5.0in]{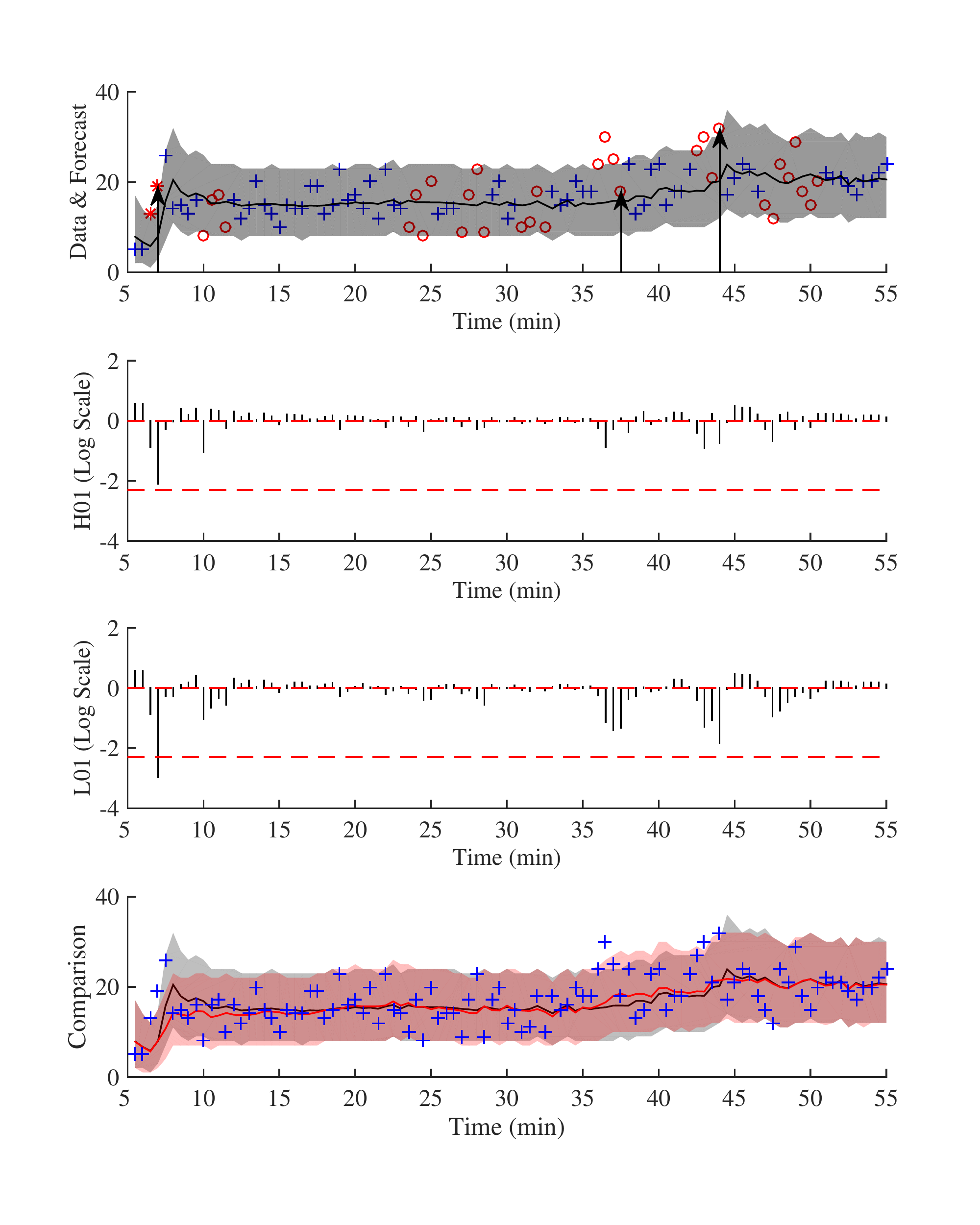}
 \caption{\small  Summaries from analysis of transitions from
node $i$=Homepage $\rightarrow j$=Science under monitoring and 
 intervention, with details as in Figure~\ref{fig:monitorHome_World}.
} 
\label{fig:monitorHome_Science}
\end{figure}

One example from the Fox News study is summarized in
Figure~\ref{fig:monitorHome_World}. While a rather extreme case in terms of one series of 
time points where the departure from the steady random-walk evolution of the BFDM is very 
apparent,  this example nicely highlights the efficacy of the 
on-line monitoring and automated intervention strategy. The example is transition flows from 
Homepage to World over the February 23rd am period. 
There are several periods in which $l_t>1$ but the evidence against the normal model is
not so strong as to signal an exception and call for intervention. The period around 23--25minutes
saw a substantial upswing in flows that triggered interventions to adapt three times.  Interventions
at about 32.5 and 48minutes  were triggered by a cumulated run-length $l_t$ suggesting 
gradual drift from the standard model. 
We also note that this picture is very similar when shown in terms of the flow 
frequencies $x_{ijt}/n_{i,t-1}$ and conditional transition probabilities $\theta_{ijt}$ rather than
raw counts $x_{ijt}$ and rates $\phi_{ijt}.$  This is an example where there were (at least) two periods
of change in transition characteristics beyond those defined by the BDFM,  but that monitoring and
intervention is able to adapt to on-line. In a real-life, sequential context, much more can and should be done, of course, at times of intervention. The analysis summary here simply serves to show the potential,
recognizing that this is applied in parallel across all inflow and node-node transition models in 
a wholly automated manner.  A second example in Figure~\ref{fig:monitorHome_Science} shows a
somewhat more typical stable trajectory, with only two interventions that appropriately adapt to 
the modest and subtle level changes in the latter part of the time period. 
 
\section{Closing Comments}\label{sec:closingcomments}

The BDFM framework is adaptive to time-varying rates of flows within 
dynamic networks and able to coherently quantify non-stationary changes in
within- and into-/out of- network flow rate processes.  This extends and
customizes non-stationary process models for count data for transition flows in a network.  
Novelties include
use of relevant occupancy factors to appropriately scale Poisson rates in sets of 
decoupled models, and the introduction of discount factor scheduling to 
appropriately address problems with, in particular, low flow rates. 
Sequential analysis of the resulting Bayesian dynamic flow model is fast and efficient. 
The decoupled analyses yield full posterior 
distributions for rate parameter process parameters across nodes
and pairs of nodes in a scalable manner.  
Our analysis of the Fox News network 
time series data sets shows the utility  of the BDFM in generating initial inferences on flow 
rate processes, in highlighting differences across days and in generating potential practical \lq\lq leads". 
On the latter, for example,  it is immediately clear from the BDFM results that 
most visitors go to just one domain, rather than
traversing to multiple domains. This has potential decision implications for
computational advertising, and also likely highlights a 
difference between on-line news consumers
and traditional newspaper readers.   
 
The emulation \lq\lq map"  from the BDFM to the DGM
represents a strategy of increasing interest in various
areas, especially with regard to efficient computation and scalability. We fit a flexible, adaptive model
in a set of decoupled analyses, and then directly map posterior samples to 
the more substantively interesting   parameter processes in a model 
that is otherwise challenging to fit.  
Applied to the Fox News flow data,  we
show some (of the many) examples of how this isolates 
dynamics in node-specific and interaction effects.  In some such cases, this 
indicates \lq\lq time-varying sparsity" in node-node interaction
effects over time, highlighted with our use of Bayesian credible values over time: 
some interaction effects (affinities)
appear significant at some points in time but not in others. A number
of the specific   node-node inferences mentioned in the application section highlight
additional results   of substantive interest, some of which are initially unexpected.
Others include a sustained positive affinity of Opinion for Homepage, but a
similarly sustained but negative affinity of Science for Homepage.   
Additionally, comparisons across different times  
of the day and across days identified and quantified patterns related to anomalous
flows corresponding to identifiable news events that appear to 
have driven traffic to specific nodes on the Fox News site.

Computational demands for the full analysis scale 
as $T(I+1)^2 n$ where $n$ is the Monte Carlo sample size. Analysis is very fast, based on the
core modeling and emulation strategy.  A 2016 Matlab implementation running on a 
standard laptop (2.3GHz cpu, 16Gb memory) took less that 5minutes to run 
our one period context with $T=110, I=22$ and $n=5{,}000.$ 
One current interest is in developing this in analysis of a more elaborate, sub-domain network
of more than 1{,}000 nodes, currently under development. 

Additional methodological development concerns the use of formal sequential 
Bayesian model monitoring based on Bayes factor tests, and the accompanying 
automated intervention analysis to allow models
to adapt at time of potential change in underlying flow parameter processes beyond 
normal levels of variation. 
Importantly, our use of decoupled/recouple models for within-network transitions 
allows statistically and computationally efficient development of sequential Bayesian 
testing based on Bayesian factors related to short-term prediction of each of the 
individual node-node flows in parallel. At each time point, the decoupled models 
are monitored, and any signals of significant departure from predictions may be 
linked across node pairs to explore for dependencies. An unusually significant 
decrease in inflow to Entertainment at one time, for example, may come via increased
transitions from Homepage alone, or represent an Entertainment effect evident in 
flows from other domains as well.  Monitoring and 
change detection in the decoupled BDFMs can lead to intervention to modify posteriors  
one or more of the decoupled posteriors for the $\phi_{ijt},$  but the mapping to 
the DGM parameters will then quantify and highlight the potential relevance for dependencies
as well as interaction/affinity effects.

There are now opportunities to use and develop these models as a basis to
characterize the stochastic dynamics of website flows, and hence feed into 
modeling and decision analysis that addresses the needs to respond to changing 
patterns in computational advertising.  An ability to rapidly signal potential 
anomalies in a small subset of domains in real-time will be of huge interest in this field.   
More immediately, some of the evident questions arising from the current study 
concern the overlay of the \lq\lq unbiased" inferences about changes and structure 
in network flows with substantive covariate information.
In many applications, including computational advertising but
also capital and transportation flows, there are useful covariates
that could inform the analysis. Our perspective here has been mainly exploratory, 
aiming to define a formal basis for effectively characterizing non-stationary
stochastic dynamics in flow data, and adapting models over time. 
One next step is to overlay any particular application
with covariate information as descriptive/explanatory as we exemplified with some
vignettes from the Fox News study.  One aspect of this is to consider random 
effects representing otherwise unexplained extra-Poisson variation that seems relevant in
some cases. A more predictive level would involve extensions 
 to incorporate covariates in the BDFMs, so future research in that direction is warranted.

Finally, we note that the general context of time-varying traffic flow analysis 
arises in multiple other fields.  Beyond in origin-destination analysis and related goals in 
studies of transportation networks 
~\citep[e.g.][]{Tebaldi1998} and physical traffic~\citep[e.g.][]{Tebaldi2002, Queen2009, AnacletoEtAl2013a, AnacletoEtAl2013b}, 
such data in neural spike train experiments, other varieties of internet 
traffic, and network studies in areas as diverse as biological anthropology (e.g., 
grooming interactions in primate troops), human social networks, flows between 
institutions in finance and economic networks, and others. 
Our work here represents new methodology  
of Bayesian dynamic modeling in an application with at least conceptual intersections with 
some of these areas, and may well be explored in such applications.

\newpage
\appendix 

\section{Appendix: Gamma-Beta Discount Models \label{app:betagammadm} }

This Appendix provides additional details and discussion of the gamma-beta
``steady"  dynamic model for time-varying Poisson rates, 
extending the background details underlying the core model summarized in 
Section~\ref{sec:gammabetadm}. 

Using generic notation, a series of non-negative counts $x_t$ over
$t=1{:}T$ is modeled via $x_t |\phi_t \sim Poi( m_t\phi_t)$ conditionally 
independently over time, where the underlying/latent Poisson rate process
$\phi_t$ follows a gamma-beta stochastic model and each $m_t$ is a
 scaling constant known at time $t.$ 
This is effectively a non-stationary, non-Gaussian random walk model, 
so it has enormous flexibility in adapting to changes over time. 
The extent of anticipated stochastic change over time is defined 
by a discount factor parameter $\delta_t\in (0,1)$, potentially 
different at each time $t$.     
We detail the model concept and structure, and the implied machinery for 
Bayesian learning and forecasting that includes the forward 
filtering, backward sampling (FFBS) algorithm for conditionally 
Poisson time series coupled with the gamma-beta steady process model.

\subsection{Forward Filtering (FF)} 
At time $t=0,$ introduce an hypothetical latent state $\phi_0$ 
and use $x_0$ as a synthetic notation for all available initial 
information. 
Specify an initial gamma prior, so $\phi_0\sim Ga(r_0,c_0)$ where
$r_0>0$, $c_0>0$ are known. 

For each $t=1{:}T,$ the model and forward/sequential analysis are then
as follows.   

\paragraph{Posterior at time $t-1$:} 
Standing at time $t-1$, the posterior for the current Poisson rate
given the initial information and all data observed over past 
times $0{:}t-1$ is gamma,   
\begin{equation}\label{eq:gammaposttm1}  
\phi_{t-1}| x_{0{:}t-1} \sim Ga(r_{t-1},c_{t-1})
\end{equation} 
where the defining parameters are known, evaluated from past
information $x_{0{:}t-1}.$

\paragraph{Evolution to time $t$:} 
The Poisson rate evolves to time $t$ via the gamma-beta evolution 
\begin{equation}\label{eq:gammabetadynamics} 
\phi_t=\phi_{t-1} \eta_t/\delta_t, \qquad   \eta_t \sim Be( \delta_t r_{t-1}, (1-\delta_t)r_{t-1}),
\end{equation}
where the random ``shock", or innovation, $\eta_t$ is independent of 
$\phi_{t-1}.$  
This is a multiplicative random walk model in that 
$E(\phi_t|\phi_{t-1}) = \phi_{t-1},$  hence the use of the 
``steady model" terminology. 
A lower value of $\delta_t$ leads to a more diffuse beta innovation 
distribution and the ability to adapt to changing rates over time, 
while a value closer to one indicates a steady, stable evolution. 
The random walk nature of the model allows for changes, 
but does not anticipate specific directional changes. 
The model results in a fully Bayesian solution to rather simple, 
flexible smoothing of discrete time series in the context of 
variation in the underlying latent process. 

 Note that the beta innovations distribution for $\eta_t$ 
at time $t$ depends in the accumulated information content about the time $t-1$ 
level through the shape parameter $r_{t-1}.$   The discount factor $\delta_t$ acts to 
decrease the information content between times $t-1$ and $t$ in a natural way.  
That is, information loss rates are constant over time, rather than parameters of the 
innovation distribution. 
The specific choice of beta distribution ensures that the implied time $t$ prior 
has a conjugate gamma form. 

\paragraph{Prior for time $t$:} 
The time $t-1$ gamma posterior of~\eqno{gammaposttm1} couples with the
beta innovation to give the time $t-1$ prior for the next state as 
\begin{equation}\label{eq:gammapriort}  
\phi_t|x_{0{:}t-1} \sim Ga(\delta_t r_{t-1},\delta_t c_{t-1}).
\end{equation}
Here we see the discounting effect of the random walk model:  
the prior for the evolved rate is more diffuse than the time 
$t-1$ posterior, reflecting increased uncertainty due to evolution.  

\paragraph{One-step ahead predictions:} 
Predicting the data at time $t,$ the one-step ahead forecast distribution is 
generalized negative binomial with p.d.f.
\begin{equation} \label{eq:onestepnegbinpdf} 
p(x_t|x_{0{:}t-1},\delta_{1{:}t})= 
	\frac{\Gamma(\delta_t r_{t-1}+x_t)}{\Gamma(\delta_t r_{t-1})\Gamma(x_t+1)} 
	\frac{m_t^{x_t} (\delta_t c_{t-1})^{\delta_t r_{t-1}}}
			{(\delta_t c_{t-1}+m_t)^{\delta r_{t-1} + x_t}}
\end{equation} 
on $ x_t=0,1,\cdots. $

\paragraph{Posterior at time $t$:} 
Observing $x_t$, the resulting posterior is 
$\phi_t|x_{0{:}t} \sim Ga(r_t, c_t)$, which has the same 
form as that at time $t-1$ but with updated  parameters  
$r_t=\delta_t r_{t-1} + x_t$ and $c_t=\delta_t c_{t-1}+m_t$.

\subsection{Model Marginal Likelihood (MML)} 
A key ingredient of formal model assessment is the model marginal
likelihood that, in this first-order Markov model, is computed as 
the product of one-step forecast p.d.f.s evaluated at the realized data.
At time $t$, this product is   
$$ 
p(x_{1{:}t}|x_0,\delta_{1{:}t}) = \prod_{s=1{:}t} p(x_s|x_{0{:}s-1},\delta_s). 
$$ 
The product is most usefully written in its one-step updated form 
\begin{equation}\label{eq:modelimml} 
p(x_{1{:}t}|x_0,\delta_{1{:}t}) = p(x_t|x_{0{:}t-1},\delta_{1{:}t}) p(x_{1{:}t-1}|x_0,\delta_{1{:}t-1})
\end{equation} 
where the contribution at time $t$ derives from the one-step ahead 
predictive density of \eqno{onestepnegbinpdf} evaluated at the 
datum $x_t.$ 
These are trivially computed. 
  
One of the most useful roles of the marginal likelihood is in comparing models based on
different (sets of) discount factor values.  As one key special case, suppose $\delta_t = \delta$ is
fixed over the time period of interest.  Then~\eqno{modelimml} gives the 
value of the marginal likelihood $p(x_{1{:}t}|x_0,\delta)$ at any chosen value of $\delta.$ 
In parallel analyses using a discrete set of $\delta$ values, the
log of the marginal likelihood is linearly accumulated as data are
sequentially processed. 
At any time $t$ this can be mapped to a posterior  
$p(\delta| x_{0{:}t}) \propto  p(\delta|x_0) p(x_{1{:}t}|x_0,\delta)$ 
and then normalized over the grid of values for inference on 
$\delta$ at any time of interest. 
This can be used to identify/choose a modal value for inference 
on the $\phi_t$ conditional on a chosen $\delta,$ or for model averaging. 

The sequentially computed contributions to the marginal 
likelihood---the realized p.d.f. ordinates 
$ p(x_t|x_{0{:}t-1},\delta)$---can be monitored sequentially 
over time to provide an on-line tracking of model performance, 
with potential uses in flagging anomalous data at one node or any 
subset of nodes, using standard Bayesian model monitoring 
concepts; see~\citet{West1986a}, 
\nocite{West1986a,West1986,West1989}
West and Harrison (1986, 1989 and Chapter 11 of 1997), 
and~\citet[][Section 4.3.8]{PradoWest2010}.
 
\subsection{Backward Sampling (BS)\label{sec:bs}} 
Reaching the end time $T$, we look back over time and revise the
summary posterior distributions for the full trajectory of the 
latent gamma process $\phi_{1{:}T}$ based on all the observed data.
This uses backward sampling based on theory in~\citet[][Section 10.8]{West1989book};
see also~\citet[][Section 4.3.7 and problem 4 of Section 4.6]{PradoWest2010}.

\begin{itemize}
\item
Sample the final rate from the time $T$ posterior
$\phi_T|x_{0{:}T},\delta_{1{:}T} \sim Ga(r_T,c_T)$.  
\item
Recurse back over time $t=T-1,T-2,\ldots,1$, at each stage 
sampling $\phi_t$ from the implied  $p(\phi _t|\phi _{t+1:T},x_{0{:}T},\delta_{1:T})$ via 
$\phi_t =  \delta_{t+1}\phi_{t+1}+\epsilon_t$ with a ``backward
innovation" $\epsilon_t$ drawn from 
$\epsilon_{it}\sim Ga((1-\delta_{t+1})r_t,c_t)$, independently of $\phi_{t+1}.$ 
\end{itemize}
Repeating the backward sampling generates a Monte Carlo sample 
of the trajectory $\phi_{1{:}T}$ from the full posterior 
$p(\phi_{1{:}T}|x_{0{:}T},\delta_{1{:}T})$ for summary inferences.

\section{Appendix B: Bayesian Model Monitoring and Adaptation \label{app:monitoring}}

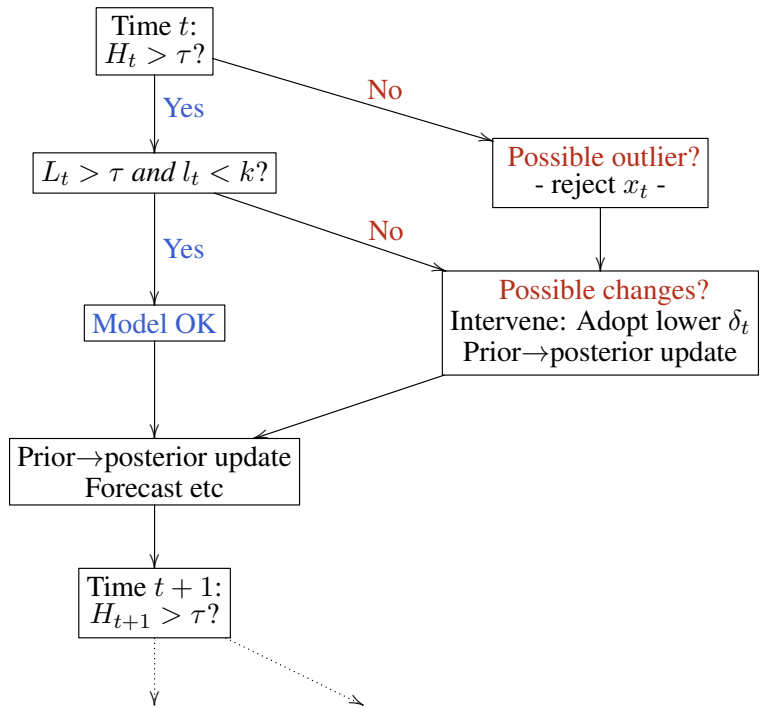
\begin{figure}[h!]
$$ 
\xymatrix{ 
	 *+[F]{\txt{Time $t$:\\ $H_t>\tau$?}} \ar[d]^{\txt{\blu{Yes}}} \ar[rrd]^{\txt{\red{No}}}
		&& 						\\
	*+[F]{\txt{$L_t>\tau$ {\em and} $l_t<k$?}}
		\ar[d]^{\txt{\blu{Yes}}} \ar[rrd]^{\txt{\red{No}}} 	
				    && *+[F]{\txt{ \red{Possible outlier?} \\ - reject $x_t$ - }}\ar[d]		 \\
	*+[F]{\txt{\blu{Model OK}}} \ar[d] 
				    && *+[F]{\txt{ \red{Possible changes?} \\ Intervene: Adopt lower $\delta_t$ 														\\ Prior$\to$posterior update}}\ar[lld]	 \\
	*+[F]{\txt{Prior$\to$posterior update\\ Forecast etc}} \ar[d] && \\ 
	*+[F]{\txt{Time $t+1$: \\ $H_{t+1}>\tau$?}} \ar@{.>}[d]\ar@{.>}[rd] && \\
	&&\\ 
}
$$ 
\caption{Schematic of monitoring, routine outlier/change assessment and automatic intervention 
to allow for more adaptability in times of change, as discussed and detailed in Section~\ref{sec:monitor}. 
This relies on a Bayes factor threshold $\tau$ (e.g. $\tau=0.1$) 
and run-length threshold  $r$ (e.g. $r=4).$     This schematic is a modified version of Figure 11.9
in~\citet{WestHarrison1997}.}
\label{fig:flowchart}
\end{figure}

%
%

\newpage


\end{document}